\documentclass[aps,prd,12pt,nofootinbib,superscriptaddress]{revtex4-2}
\DeclareRobustCommand{\baselinestretch{1.27}}

\usepackage{amsmath,amssymb,amsbsy}
\usepackage{graphicx}
\usepackage{subfigure}

\def\bx{{\boldsymbol x}}
\def\by{{\boldsymbol y}}
\def\bk{{\boldsymbol k}}
\def\bp{{\boldsymbol p}}
\def\bl{{\boldsymbol l}}
\def\bn{{\boldsymbol n}}
\def\bv{{\boldsymbol v}}

\def\cP{{\cal P}}

\def\ri{{\rm i}}

\usepackage{hyperref}
\hypersetup{
	colorlinks=true,
	allcolors=blue,
	linktocpage=true,
	pdfhighlight=/N
}

\begin{document}

\title{Discreteness effects in $N$-body simulations \\ of warm dark matter}

\author{Yuri Shtanov}
\email{shtanov@bitp.kyiv.ua}
\affiliation{Bogolyubov Institute for Theoretical Physics,  Metrologichna Street~14-b, Kyiv 03143, Ukraine} %
\affiliation{Astronomical Observatory, Taras Shevchenko National University of Kyiv, Observatorna Street~3, Kyiv 04053, Ukraine} %

\author{Valery I. Zhdanov}
\email{valery.zhdanov@knu.ua}
\affiliation{Astronomical Observatory, Taras Shevchenko National University of Kyiv,  Observatorna Street~3, Kyiv 04053, Ukraine} %

\begin{abstract}
In cosmological $N$-body simulations of warm dark matter, thermal velocities of dark-matter particles are sometimes taken into account by adding random initial velocities to the particles of simulation. However, a particle in the $N$-body system represents a huge collection of dark-matter particles, whose average thermal velocity is very close to zero. We consider justification of the procedure of adding thermal velocities in $N$-body simulations and build a simple model of their influence on the power spectrum. Our model captures the physical effect of suppression of the power spectrum at small wave numbers and also explains its artificial enhancement at large wave numbers, observed in numerical simulations with added thermal velocities. The cause of this enhancement is the disturbance of the growth rate of the density profile introduced when adding random initial thermal velocities. Specifically, the model predicts a turnover in the behavior of the simulated power spectrum at a certain wave number $k_*$, beyond which it grows as $P (k) \propto k^2$. Our treatment is generalized to a system consisting of several matter components with different thermal velocity dispersion. We also estimate the effects of discreteness related to the bulk velocity field and establish the conditions under which these effects dominate over those of thermal velocities.
\end{abstract}

\maketitle

\section{Introduction}

Comparisons between theory and observations indicate the cosmological presence of dark matter (DM), which is widely believed to consist of nonbaryonic particles. The nature of these particles is unknown, and their basic parameters can vary in a considerably wide range, affecting their production mechanisms and evolution history. One of the important characteristics of dark matter is its particle velocity distribution.  Dark matter that decouples (both chemically and kinetically) from thermal equilibrium in the primeval plasma deeply at the radiation-dominated epoch while being nonrelativistic has a Boltzmann velocity distribution function at the moment of freeze-out and is dubbed as ``cold dark matter'' (CDM)\@. In the CDM scenario, primordial density perturbations grow on all spatial scales of interest in cosmology, and structure formation proceeds in the bottom-up fashion: the smallest structures form first, subsequently combining into larger structures.

Dark matter which is nonrelativistic at the time of matter--radiation equality ($z = z_\text{eq} \simeq 3 \times 10^3$) but whose particle velocity distribution cannot be neglected in the theory of structure formation is called ``warm dark matter'' (WDM)\@.  A typical example is super-weakly interacting particles that are produced deeply at the radiation-dominated epoch while being relativistic. Density perturbations of WDM particles are suppressed on spatial scales below their free-streaming length. The largest spatial scale affected by free streaming is determined by the distance covered by a dark-matter particle moving with a typical velocity from the moment of its production or decoupling.

The form of the velocity distribution function of dark-matter particles depends on the mechanism of their production or decoupling. A typical example is primordial phase-space distribution function of the form\footnote{Our general consideration below is carried out at a classical level and is not limited to a specific type of particles (bosons or fermions) or their specific velocity distribution. Equation \eqref{fermi} is used to illustrate one of the possibilities.}
\begin{equation} \label{fermi}
f(p) = \frac{\chi}{\exp( p / T_\text{DM}) + 1 } \, ,
\end{equation}
where $p$ is the particle momentum, $T_\text{DM}$ is the effective temperature, and $\chi \leq 1$ is the normalization factor.  Such a distribution function arises in the case of thermal relics that are highly relativistic at the time of decoupling; in this case, $\chi = 1$, and the temperature $T_\text{DM} = T_\text{TR}$ is deduced from the mass and abundance of these particles. A similar distribution arises also in the scenario of nonresonantly produced sterile neutrinos \cite{Dodelson:1993je, Dolgov:2000ew} (see \cite{Boyarsky:2018tvu} for a review), in which case $\chi < 1$, and, typically, $T_\text{DM} = T_\nu$, the temperature of active neutrinos. In view of the thermal nature of distribution \eqref{fermi}, one usually refers to the correspondingly distributed velocities of dark-matter particles as to {\em thermal velocities\/}. One can use this term also in more general situations, in which the particle velocity distribution does not have thermal character still featuring considerable dispersion.

Discreteness effects in numerical cosmological simulations of cold dark matter are the subject of numerous investigations (which are usually also done numerically, see, e.g., \cite{Romeo:2008jx}). In this paper, we consider, purely analytically, the issue of discreteness effects in numerical simulations of dark matter with thermal velocities. In the majority of papers on simulations of the cosmological large-scale structure formation, the ``warmth'' of dark matter is taken into account only in its resulting free-streaming effect of suppression of the power spectrum on small spatial scales, while the particles themselves are treated as cold in the initial conditions. The justification of this lies in the observation that thermal velocities of WDM particles typically become small compared to their regular (bulk) velocities by the time when structure formation starts to become nonlinear. Even in this case, one observes discreteness effects on small spatial (or mass) scales, such as formation of a large number of spurious clumps resembling dark-matter halos \cite{Wang:2007he}. Recently, new numerical methods were developed to simulate cosmologies with suppression of density perturbations on small spatial scales that avoid these numerical artifacts \cite{Hahn:2015sia, Sousbie:2015uja, Stucker:2019txm, Stucker:2021vyx}. The core of these methods is based on the fact that cold dark-matter particles occupy a three-dimensional surface in a six-dimensional phase space, which initially has relatively smooth shape.

However, in some situations, thermal velocities of particles themselves may be non-negligible, making the simulation problem essentially six-dimensional in phase space. This is the case, e.g., when performing simulations of warm dark matter starting from high cosmological redshifts or when a fraction of dark matter is composed of a light thermal relic (which can be the usual neutrino).  In such cases, initial random thermal velocities are usually added to the simulation particles \cite{Klypin:1992sf, Colin:2007bk, Brandbyge:2008rv, Viel:2010bn, Leo:2017zff}. In this paper, we would like to address the issue of the discreteness effects caused by adding thermal velocities in cosmological $N$-body simulations.  

Random initial velocities are added to the simulation particles according to the initial {\em velocity\/} distribution function of the physical WDM particles. However, because of obvious computational limitations, a particle in a simulation represents matter in a large region with characteristic size $\ell \simeq \left(V / N \right)^{1/3}$ (where $V$ is the box volume and $N$ is the number of simulation particles). Therefore, a huge number of DM particles are represented as one particle in $N$-body simulations. Formally, the average thermal velocity of such a collection of DM particles is very close to zero. There arises the question about the justification of the procedure of adding thermal velocities and about possible discreteness effects that may appear in such simulations. We think it would be useful to provide a theoretical assessment of these effects, which usually are treated empirically by performing real $N$-body simulations. 

In the present paper, we build a simple theoretical model of $N$-body simulations of warm dark matter and elucidate the effects arising by adding thermal velocities in such simulations. The problem is analyzed by using the Peebles equation describing the evolution of the Fourier components of the density contrast. This equation allows for a rather simple and direct analysis of the case of finite $N$ and has a natural physical limit of $N \to \infty$.

Our analysis reveals that the main side effect of adding initial thermal velocities in simulations consists in perturbing the initial growth rate of the density profile at that moment. This perturbation then propagates to the future resulting in an artificial increase of power at small spatial scales.  We obtain an analytic formula for the resulting power spectrum that can be used for theoretical estimates of this effect. In the limit of $N \to \infty$, this discreteness effect vanishes, and only the physical effect of thermal velocities remains, which consists in a natural suppression of the power spectrum.

Our theory and main results will be fairly general, describing particles with thermal velocity distribution of any form. We will see that, basically, only the dispersion of this thermal velocity is going to be important for our general results. However, in order to illustrate our results for concrete situations, as an example, we will often use distribution \eqref{fermi}, specific to sterile neutrino.

Our paper is organized as follows. In Sec.~\ref{sec:evolution}, we introduce the Peebles equation \cite{Peebles-AA, Peebles:1980yev} describing the evolution of the density contrast in a system composed of particles.  In Sec.~\ref{sec:coarse}, we describe the procedure of coarse-graining of the DM particle system, relevant to $N$-body simulations, identifying the shot noise. In Sec.~\ref{sec:model}, we build our statistical model of $N$-body simulations, deriving an approximate equation for the evolution of the density contrast of warm dark matter that captures the presence of thermal velocities.  The effects of thermal velocities in $N$-body simulations with finite box volume $V$ and particle number $N$ are then studied in Sec.~\ref{sec:noise}\@.  We demonstrate the existence of numerical artifacts that tend to increase the power spectrum; they are proportional to the inverse of particle number density $V / N$, and depend on the wave number as $k^2 / k_\text{th}^2$, where $k_\text{th}$ is the {\em thermal\/} wave number arising in the presence of thermal velocities. We obtain an analytic formula for the resulting power spectrum containing both the physical effects and the artifacts of discreteness connected with thermal velocities. In Sec.~\ref{sec:several}, we discuss extension of our analysis to the case of dark matter composed of several components with different thermal velocity distributions. In Sec.~\ref{sec:field}, we analyze the discreteness effect connected with the effective coarse-graining of the regular velocity field in $N$-body simulations.  We show that it also tends to increase the power spectrum because of the presence of the term proportional to $k^2 / k_\text{fi}^2$ in the effective evolution equation for the density contrast, where $k_\text{fi}$ is the characteristic wave number connected with the coarse-graining of the smooth velocity field.  We determine the parameter region for which this effect is more important than the physical effect of thermal velocities. In Sec.~\ref{sec:discuss}, we summarize the main results of the paper and discuss their implications.  Various definitions and technical details are presented in Appendices \ref{app:power} and \ref{app:spectra}.

\section{Preliminaries: evolution of the mass density}
\label{sec:evolution}

We consider a spatially flat universe dominated by nonrelativistic matter and cosmological constant.  This is a fairly good approximation to the later stages of cosmological evolution.

On sub-Hubble spatial scales, an instantaneous state of matter is described by the mass density $\rho (\bx)$ with mean background value $\varrho$.  It is realized by a large number of particles of (usually equal) masses $m_i$ placed at comoving coordinates $\bx_i$.  We then have, accordingly, 
\begin{equation} \label{density}
\rho (\bx) = \sum_i \frac{m_i}{a^3} \delta_\text{D} \left( \bx - \bx_i \right) \, , 
\end{equation}
where $a$ is the scale factor of the expanding universe, and $\delta_\text{D} (\bx)$ is Dirac's delta-function.

The particle equation of motion is
\begin{equation} \label{eqmot}
\ddot \bx_i + 2 H \dot \bx_i = - \frac{1}{a^2} \nabla \varphi \left( \bx_i \right) \, ,
\end{equation}
where $H \equiv \dot a / a$ is the Hubble parameter, and $\varphi (\bx )$ is the gravitational potential that satisfies the equation
\begin{equation} \label{Poisson}
\nabla^2 \varphi (\bx) = 4 \pi G a^2 \left[ \rho (\bx) - \varrho \right] \, .
\end{equation}

In this paper, we use the formalism introduced by Peebles in \cite{Peebles-AA} (see also \cite[Sec.~27]{Peebles:1980yev}), which appears to be best suited to study the effects arising in $N$-body simulations. For particles in a (periodic) box of comoving volume $V$, one can proceed to the Fourier transform of the density contrast $\delta (\bx) \equiv \left[\rho (\bx) - \varrho \right] / \varrho\,$:
\begin{equation}
\quad \delta_\bk = \frac{1}{V} \int_V \delta (\bx)  e^{\ri\bk\bx} d^3 \bx  = \left\{
\begin{array}{cl} \displaystyle\sum_i \frac{m_i}{M} e^{\ri\bk\bx_i} \, , &\bk \ne 0 \, , \medskip \\ 0 \, , &\bk = 0 \, , \end{array} \right.
\end{equation}
where $M$ is the total mass of particles, $M = \sum_i m_i$. For the gravitational potential, using \eqref{Poisson}, one then finds
\begin{equation} \label{potential}
\varphi (\bx) = \sum_\bk \varphi_\bk e^{ - \ri \bk \bx} = - 4 \pi G a^2 \varrho \sum_{\bk \ne 0} e^{- \ri \bk \bx} \frac{\delta_\bk}{k^2} \, .
\end{equation}
Note that $\varphi_{\bk=0} = 0$ by definition. 

After that, using \eqref{eqmot}, one can derive a very useful equation describing the temporal evolution of the Fourier amplitudes $\delta_\bk$ \cite{Peebles-AA, Peebles:1980yev}:  
\begin{equation} \label{nonlin}
\ddot \delta_\bk + 2 H \dot \delta_\bk = 4 \pi G \varrho \delta_\bk + A_\bk - C_\bk \, ,
\end{equation}
where
\begin{align}\label{A}
A_\bk &= 4 \pi G \varrho \sum_{\bk' \not\in \{0, \bk\}} \frac{\bk \bk'}{k'^2} \delta_{\bk - \bk'} \delta_{\bk'} \, , \\
C_\bk &= \sum_i \frac{m_i}{M} \left( \bk \dot \bx_i \right)^2 e^{\ri \bk \bx_i} \, .
\label{C}
\end{align}

We note that Eq.~\eqref{nonlin} is an exact equation for the evolution of the density profile, valid in the case of universe composed of the matter under consideration and cosmological constant. It is applicable both to real particles in the universe and to particles in $N$-body simulations. This equation is not closed with respect to $\delta_\bk$ because of the term $C_\bk$, which involves particle velocities, and thus requires an equation describing the evolution of the velocity field (see \cite[Sec.~27]{Peebles:1980yev}).

\section{Coarse-graining and thermal velocities}
\label{sec:coarse}

Numerical simulations deal with a coarse-grained system, in which the particle mass exceeds the mass of a fundamental dark-matter particle by many orders of magnitude. For example, in \cite{Leo:2017zff}, the simulation particle mass is around $10^7\, h^{-1} M_\odot$, while the dark-matter particle mass is of the order of keV, i.e., $10^{-63} h^{-1} M_\odot$. So each simulation particle represents about  $10^{70}$ physical particles. In this section, we briefly describe this coarse-graining and its effects.

Regarding distribution \eqref{density} as the distribution of real particles in the universe, we form its coarse-grained realization as follows: the coordinate space is partitioned into $N$ regions, numbered by $I$, with (equal, for simplicity) total mass $m_I$ within each region, and original particles of small mass $m_i$ within each region are collected {\em to the center of mass\/} of the corresponding region to form a particle of mass $m_I$. For brevity, we will call the collection of original particles in the $I$th region as the $I$th cluster. We get a coarse-grained distribution and its Fourier transform, which we denote by an overbar:
\begin{equation}
\bar \rho (\bx) = \sum_I \frac{m_I}{a^3} \delta_\text{D} \left( \bx - \bx_I \right) \, , \qquad \bar \delta_\bk = \int \frac{ \bar \rho (\bx) - \varrho}{\varrho V} e^{\ri\bk\bx} d^3 \bx  = \sum_I \frac{m_I}{M} e^{\ri\bk\bx_I} \, , \quad \bk \ne 0 \, .
\end{equation}

There are several effects connected with coarse-graining. First of all, we would like to estimate the difference between $\bar \delta_\bk$ and $\delta_\bk$ at wavelengths much larger than the mean comoving separation between the masses $m_I$. This is done by noting that the position $\bx_{i_I}$ of each particle in the $I$th cluster is obtained by a shift $\by_{i_I}$ from the center of mass $\bx_I$ of the cluster: $\bx_{i_I} = \bx_I + \by_{i_I}$. Therefore, one can write
\begin{align}\label{relation}
\delta_\bk &= \sum_i \frac{m_i}{M} e^{\ri\bk\bx_i} = \sum_I \frac{m_I}{M} e^{\ri\bk\bx_I} \sum_{i_I} \frac{m_{i_I}}{m_I} e^{\ri\bk\by_{i_I}} \nonumber \\
&= \sum_I \frac{m_I}{M} e^{\ri\bk\bx_I} \sum_{i_I} \frac{m_{i_I}}{m_I} \left[ 1 + \ri\bk\by_{i_I} - \frac12 \left( \bk\by_{i_I} \right)^2 + \ldots \right] \nonumber \\
&= \sum_I \frac{m_I}{M} e^{\ri\bk\bx_I} \sum_{i_I} \frac{m_{i_I}}{m_I} \left[ 1 - \frac12 \left( \bk\by_{i_I} \right)^2 + \ldots \right] \, , 
\end{align}
where we have expanded the second exponent according to our condition $k y_i \ll 1$. The last equality in \eqref{relation} follows from the center-of-mass property $\sum_{i_I} m_{i_I} \by_{i_I} = 0$. Introducing the notation
\begin{equation} \label{eln}
\ell_I^2  (\bn) \equiv \sum_{i_I} \frac{m_{i_I}}{m_I} \left( \bn \by_{i_I}
\right)^2 \, , \qquad \bn = \frac{\bk}{k} \, ,
\end{equation}
and neglecting higher powers of $k y_i$, we have
\begin{align} \label{rel}
\delta_\bk &\simeq \sum_I \frac{m_I}{M} e^{\ri\bk\bx_I} \left[ 1 - \frac12 k^2 \ell_I^2 (\bn) \right] \nonumber \\ &= \sum_I \frac{m_I}{M} e^{\ri\bk\bx_I} \left( 1 - \frac12 k^2 \ell^2 \right) + \frac12 \sum_I \frac{m_I}{M} e^{\ri\bk\bx_I} k^2 \left[ \ell_I^2 (\bn) - \ell^2 \right] \nonumber \\ 
&= \bar \delta_\bk \left[ 1 - \frac12 ( k \ell )^2 \right] + \frac12 \sum_I \frac{m_I}{M} e^{\ri\bk\bx_I} k^2 \left[ \ell_I^2 (\bn) - \ell^2 \right] \, ,
\end{align}
where $\ell^2$ is the average value of $\ell_I^2 (\bn)$ over $I$. The last term in this expression is the sum of $N$ random walks with characteristic step $\sim k^2 \ell^2 / N$. It produces the shot noise in the squared amplitude with the characteristic level $\left| \delta_\bk \right|^2 \sim k^4 \ell^4 / N$ \cite{Peebles:1980yev, Zeldovich:1965gev}.

The length $\ell$ is similar to the inter-cluster distance, i.e., the distance between particle in simulations, hence, $\ell \simeq \left( V / N \right)^{1/3}$. Introducing the Nyquist wave number $k_N$ by
\begin{equation} \label{kN}
k_N = \frac{\pi}{\ell} = \pi \left(\frac{N}{V} \right)^{1/3} \, ,
\end{equation}
from \eqref{rel} we obtain a relation between the coarse-grained and fine-grained power spectra:\footnote{Definitions of the power spectrum are presented in Appendix~\ref{app:power}.}
\begin{equation}\label{coarse}
\bar P(k) \simeq P(k) \left( 1 + \frac{\pi^2 k^2}{k_N^2} \right) + P_N (k) \, , \quad k \ll k_N \, ,
\end{equation}
where
\begin{equation}\label{shot-l}
P_N (k) \simeq \frac{V}{N} \left( \frac{\pi k}{k_N} \right)^4 \, , \quad k \ll k_N \, .
\end{equation}

Thus, coarse-graining modifies the power on large scales producing relative noise of magnitude $P (k) (\pi k / k_N)^2$ as well as a shot noise $P_N (k)$.  Perhaps, if necessary, the relative noise can be corrected in simulations by modifying the input power spectrum $P (k)$ to compensate for the numerical factor in the first term of \eqref{coarse}.  

For wave numbers $k$ exceeding the Nyquist wave number \eqref{kN}, the power spectrum is dominated by the shot noise, which can be obtained by averaging over random realizations of the configurations $\{ \bx_I \}\,$:
\begin{equation}\label{shot-s}
P_N (k) = V \overline{\left| \delta_\bk \right|^2 } = V \overline{\left| \frac{1}{N} \sum_I e^{\ri\bk\bx_I} \right|^2} = \frac{V}{N} \, , \qquad k \gtrsim k_N \, .
\end{equation}
We will use an interpolating formula capturing the two asymptotics  \eqref{shot-l} and \eqref{shot-s} for the shot noise for all values of $k$:
\begin{equation}\label{shot}
P_N (k) = \frac{V}{N} \left[ 1 + \left( \frac{k_N}{\pi k} \right)^4 \right]^{-1} \, .
\end{equation}

Thus far, we were discussing the effects of coarse-graining on the Fourier transform $\delta_\bk$ of the density profile and on the related power spectrum. However, the evolution equation \eqref{nonlin} also involves the term $C_\bk$ dependent on the particle velocities. Therefore, it is necessary to consider the properties of this term under coarse-graining. By definition, in this procedure, we have
\begin{align} \label{cc}
C_\bk &= \sum_i \frac{m_i}{M} \left( \bk \dot \bx_i \right)^2 e^{ \ri \bk \bx_i} = \sum_I e^{\ri \bk \bx_I} \sum_{i_I} \frac{m_{i_I}}{M} \left[ \bk \left( \dot \bx_I + \dot \by_{i_I} \right) \right]^2 e^{\ri \bk \by_{i_I}} \nonumber \\
&= \sum_I e^{\ri \bk \bx_I} \sum_{i_I} \frac{m_{i_I}}{M} \left[ \left( \bk \dot \bx_I \right)^2 + 2 \left( \bk \dot \bx_I \right) \left( \bk \dot \by_{i_I} \right) + \left( \bk \dot \by_{i_I} \right)^2 \right] \left[ 1 + \ri \bk \by_{i_I} - \frac12 (\bk \by_{i_I})^2 + \ldots \right]
\nonumber \\
&\approx \sum_I \frac{m_I}{M} \left( \bk \dot \bx_I \right)^2 e^{\ri \bk \bx_I} + \sum_I \frac{m_I}{M} e^{\ri \bk \bx_I} \sum_{i_I} \frac{m_{i_I}}{m_I} \left( \bk \dot \by_{i_I} \right)^2 + {\cal O} \left[ \left( k v \right)^2 k \ell \right] \, . 
\end{align}
The second term in the last expression in \eqref{cc} is a contribution from the internal velocities of particles within clusters.  The velocity $\dot \by_{i_I}$ of a particle can be split into two parts: the ``field'' part $\dot \by_{i_I}^\text{fi}$ due to the inhomogeneity of the smooth original velocity field and the ``thermal'' part $\dot \by_{i_I}^\text{th}$ due to random thermal particle velocities.  Assuming that these two parts are uncorrelated within a cluster, we have $\dot \by_{i_I} = \dot \by_{i_I}^\text{fi} + \dot \by_{i_I}^\text{th}$ and
\begin{equation}
\sum_{i_I} \frac{m_{i_I}}{m_I} \left( \bk \dot \by_{i_I} \right)^2 = \sum_{i_I} \frac{m_{i_I}}{m_I} \left[ \left( \bk \dot \by_{i_I}^\text{fi} \right)^2 + \left( \bk \dot \by_{i_I}^\text{th} \right)^2 \right] \, .
\end{equation}
Thus, eventually, we have
\begin{equation} \label{ccc}
C_\bk \approx \bar C_\bk + C_\bk^\text{fi} + C_\bk^\text{th} \, ,
\end{equation}
where
\begin{equation} \label{ccc1}
\bar C_\bk = \sum_I \frac{m_I}{M} \left( \bk \dot \bx_I \right)^2 e^{\ri \bk \bx_I} \, , \quad C_\bk^\text{fi} = \sum_I \frac{m_I}{M} \left( \bk \dot \by^\text{fi} \right)_I^2 e^{\ri \bk \bx_I} \, , \quad C_\bk^\text{th} = \sum_I \frac{m_I}{M} \left( \bk \dot \by^\text{th} \right)_I^2 e^{\ri \bk \bx_I} \, ,
\end{equation}
are the coarse-grained part, field part, and thermal part, respectively, and our notation is
\begin{equation} \label{ccc2}
\left( \bk \dot \by \right)_I^2 \equiv \sum_{i_I} \frac{m_{i_I}}{m_I} \left( \bk \dot \by_{i_I} \right)^2 \, .
\end{equation}

It should be emphasized that, in a coarse-grained system (in $N$-body simulations), the terms $C_\bk^\text{fi}$ and $C_\bk^\text{th}$ are no longer present. However, their presence in Eq.~\eqref{nonlin} is required in order that this equation reproduce the correct evolution of the real particle distribution. That is, at least initially, one should ensure that the velocity term $\bar C_\bk$ for the coarse-grained distribution matches $C_\bk$ of the fine-grained (exact) distribution. In particular, if the thermal part $C_\bk^\text{th}$ dominates in decomposition \eqref{ccc}, then, in order to capture its effect, one should give the coarse-grained particles a corresponding initial velocity dispersion.  This justifies the procedure of distributing random thermal velocities directly over simulation particles in spite of the fact that they represent huge collections of mass.

We can estimate the last two parts in \eqref{ccc1} as 
\begin{equation} \label{creg}
C_\bk^\text{fi} = \left( \frac{k v_\text{fi}}{a} \right)^2 \delta_\bk \, , \qquad C_\bk^\text{th} = \left( \frac{k v_\text{th}}{a} \right)^2 \delta_\bk \, ,
\end{equation}
where $v_\text{fi}$ is the characteristic relative velocity between neighboring particles in a coarse-grained distribution arising due to inhomogeneity of the regular velocity field, and $v_\text{th}$ is the characteristic thermal velocity (these velocities will be rigorously defined below). This estimate shows the relative importance of these two terms in \eqref{ccc}.  We will postpone the effects caused by coarse-graining of the regular velocity field till Sec.~\ref{sec:field}\@.  Meanwhile, we will be interested in the effects of thermal velocities.

\section{Model of $N$-body simulations}
\label{sec:model}

From this point on, we deal with a finite system appropriate for numerical simulations, and small Latin indices $i, j, \ldots$ will label the simulation particles, running from 1 to $N$.  In the theoretical limit of very large $N$ (formally, $N \to \infty$), the system will reproduce the real physical situation.

In accordance with the reasoning of the preceding section, it is required to add random initial thermal velocities to the simulation particles if one wishes to capture their effects.  The initial peculiar velocity of a particle having position $\bx_i$ in numerical simulations can, therefore, be split into its {\em regular\/} velocity $\bv^\text{reg}_i \equiv a \dot \bx^\text{reg}_i$, which would be assigned to CDM particles (e.g., by using the Zeldovich approximation, see \cite{Klypin:2000ku}), and a random {\em thermal\/} velocity, denoted by $\bv^\text{th}_i \equiv a \dot \bx^\text{th}_i$, with zero average, which is added to particle velocities when simulating WDM\@.

The velocity term \eqref{C} provides a criterion of importance of thermal velocities in the initial conditions for warm dark matter. Assuming that thermal velocities are uncorrelated with the regular peculiar velocities (specified by the bulk motion), we have
\begin{equation}\label{ck}
C_\bk = \sum_i \frac{ m_i}{M} \left[ \bk \left( \dot \bx_i^\text{reg} + \dot \bx_i^\text{th} \right) \right]^2 e^{\ri \bk \bx_i} = C_\bk^\text{reg} + C_\bk^\text{th}  \, ,
\end{equation}
where the regular and thermal contributions to the velocity term $C_\bk$ are defined,
respectively, as
\begin{equation} \label{cregth}
C_\bk^\text{reg} = \sum_i \frac{m_i}{M} \left( \bk \dot \bx_i^\text{reg} \right)^2 e^{\ri \bk \bx_i}\, , \qquad C_\bk^\text{th} = \sum_i \frac{m_i}{M} \left( \bk \dot \bx_i^\text{th} \right)^2 e^{\ri \bk \bx_i} \, .
\end{equation}

By assumption, the quantities $\dot \bx_i^\text{th}$ are random with zero average over the ensemble of particles (or over realizations).  Assuming the isotropy of the thermal velocity distribution, it is convenient to denote by $v_\text{th}$ the one-component dispersion
\begin{equation} \label{veldisp}
v_\text{th} = a \left[ \overline{\left( \bl \dot \bx_i^\text{th} \right)^2} \right]^{1/2} = a \left[ \frac13 \overline{\left(\dot \bx_i^\text{th} \right)^2} \right]^{1/2} \, ,
\end{equation}
where $\bl$ is a fixed unit vector in any spatial direction. Taking $\bl = \bn \equiv \bk / k$, we present the thermal part in \eqref{cregth} as
\begin{align} \label{cth}
C_\bk^\text{th} &= \left( \frac{k v_\text{th}}{a} \right)^2 \sum_i \frac{m_i}{M} e^{\ri \bk \bx_i} + \sum_i \frac{m_i}{M} \left[ \left( \bk \dot \bx_i^\text{th} \right)^2 - \left( \frac{k v_\text{th}}{a} \right)^2 \right] e^{\ri \bk \bx_i} \nonumber \\ 
&= \left( \frac{k v_\text{th}}{a} \right)^2 \delta_\bk + \sum_i \frac{m_i}{M}\, \xi_i e^{\ri \bk \bx_i} \, ,
\end{align}
where
\begin{equation} \label{xi}
\xi_i = \left( \bk \dot \bx_i^\text{th} \right)^2 - \left( \frac{k v_\text{th}}{a} \right)^2
\end{equation}
are random quantities with zero average over realizations. Their dispersion is given by
\begin{equation} \label{xidis}
\overline{\xi_i^2} = c_\text{th} \left( \frac{k v_\text{th}}{a} \right)^4 \, ,
\end{equation}
where $c_\text{th}$ is a numerical constant of order unity. For distribution \eqref{fermi} it
is equal to
\begin{equation}
c_\text{th} = \left[ \frac{189\, \zeta (3) \zeta (7)}{50\, \zeta^2 (5)} - 1 \right] \approx 3.3 \, ,
\end{equation}
while, for a Gaussian distribution, we would have $c_\text{th} = 2$. Since the random quantities $\xi_i$ are independent for different $i$, the dispersion of the second term in \eqref{cth} over realizations of $\{ \xi_i \}$ can be calculated as
\begin{equation}\label{xi-fluct}
\overline{ \left| \sum_i \frac{m_i}{M} \xi_i e^{\ri \bk \bx_i} \right|^2 } = \frac{1}{N^2} \sum_i \overline{\xi_i^2} \simeq \frac{c_\text{th}}{N} \left( \frac{k v_\text{th}}{a} \right)^4 \, .
\end{equation}
In the limit of $N \to \infty$ (proceeding to a system of physical particles), the second term on the right-hand side of \eqref{cth} is, therefore, insignificant.  

Using \eqref{nonlin} and relations \eqref{ck}--\eqref{cth}, we obtain the following equation capturing the effect of thermal velocities:
\begin{equation} \label{eff}
\ddot \delta_\bk + 2 H \dot \delta_\bk = \left[ 4 \pi G \rho - \left( \frac{k v_\text{th}}{a} \right)^2 \right] \delta_\bk + A_\bk - C_\bk^\text{reg} + \sum_i \frac{m_i}{M}\, \xi_i e^{\ri \bk \bx_i} \, ,
\end{equation}
where the regular term $C_\bk^\text{reg}$ is defined in \eqref{cregth}.

The criterion of importance of the thermal-velocity term in the brackets on the right-hand side of \eqref{eff} is given by the comparison of $(k v_\text{th} / a)^2$ with $4 \pi G \varrho$, which is characterized by the thermal wave number
\begin{equation} \label{kth}
k_\text{th}  \equiv \left( \frac{4 \pi G \varrho a^2}{v_\text{th}^2} \right)^{1/2} \, .
\end{equation}
The thermal velocity $v_\text{th}$ enters Eq.~\eqref{eff} as a speed of sound, and the thermal wave number has the same expression in terms of the thermal velocity as the Jeans wave number for a nonrelativistic isothermal gas. An intrinsic difference between the two system is that the particles of gas experience collisions and can be described by an ideal-fluid approximation on spatial scales larger than their mean free path, while the dark-matter particles stream freely. For this reason, the spatial scale $2 \pi / k_\text{th}$ is sometimes called free-streaming scale (see \cite{Boyarsky:2008xj}).  

Note that if we neglect the term $A_\bk$, which is nonlinear in the density profile $\delta_\bk$, the regular-velocity term $C_\bk^\text{reg}$ and the last term in \eqref{eff}, which describes the thermal-velocity fluctuations and statistically vanishes in the continuum limit $N \to \infty$ [see \eqref{xi-fluct}], we arrive at a linear equation for density perturbations of a fluid with the effective speed of sound $c_s = v_\text{th}$. Equation of this kind naturally arises in models with analytic treatment of WDM density perturbations in linear approximation; see, e.g., \cite{Pordeus-da-Silva:2021ujr}. 

Strictly speaking, the above splitting of the peculiar velocities of simulation particles into their regular and thermal parts can be defined only at the initial time, at which the initial velocities are set. In the course of evolution, simulation particles move away from their initial positions, and it is no longer possible operationally to split their intrinsic total peculiar velocity into a regular part and a thermal part. However, one can imagine a parallel evolution of a fine-grained (physical) system with matching initial conditions, for which our system is initially a coarse-grained one. For such a fine-grained system, the corresponding {\em coarse-grained\/} velocity field is defined at any position and at any moment of time, and one can attribute a regular part of velocity to a simulation particle by using this velocity field (the remainder will then be its thermal velocity).  We note that this represents only a theoretical way of splitting a well-specified total particle peculiar velocity $\dot \bx$ into its regular and thermal parts which then results in a decomposition \eqref{ck} of the total velocity term. We assume that relations \eqref{ck}--\eqref{veldisp} involving regular and thermal velocities defined in such a manner remain to be valid in the course of evolution. In particular, we neglect possible cross-correlations between the regular and thermal parts of velocities thus defined that can arise in the course of evolution (such correlations are absent initially). This is justifiable as, practically, it is either the thermal velocity or the regular velocity that is going to dominate in the velocity term $C_\bk$. In this case, equation \eqref{eff} will also remain to be a good approximation. 

The main problem we are interested in is formulated as follows. Assume that we run two simulations with identical initial positions $\bx_i$ at some initial moment of time $t_\text{in}$, with and without initial thermal velocities. Denote the density contrast in the evolution without thermal velocities by\footnote{Note that $\delta_\bk^\text{cold}$ at $t = t_\text{in}$ may have arbitrary power spectrum (e.g., suppressed at small spatial scales due to the effects of free-streaming that took place before the time $t_\text{in}$). The superscript ``cold'' here only means that its evolution {\em from that moment on\/} is governed by the dynamics of particles without thermal velocities.} $\delta_\bk^\text{cold}$, and that with thermal velocities by $\delta_\bk^\text{warm}$.  Our main task is to estimate the effect of initial thermal velocities on the evolution of $\delta_\bk^\text{warm}$ in such simulations, as compared to the evolution of $\delta_\bk^\text{cold}$.  

To quantify this effect, we introduce the difference
\begin{equation}
\Delta \delta_\bk \equiv \delta_\bk^\text{cold} - \delta_\bk^\text{warm} \, .
\end{equation}
Taking the difference of the corresponding differential equations \eqref{eff}, we obtain an equation for this difference $\Delta \delta_\bk$:
\begin{equation} \label{exact}
\Delta \ddot \delta_\bk + 2 H \Delta \dot \delta_\bk - \left[ 4 \pi G \varrho - \left( \frac{k v_\text{th}}{a} \right)^2 \right] \Delta \delta_\bk = \left( \frac{k v_\text{th}}{a} \right)^2 \delta_\bk^\text{cold} + \Delta \left( A_\bk - C_\bk^\text{reg} \right) - \sum_i \frac{m_i}{M}\, \xi_i e^{\ri \bk \bx_i} \, .
\end{equation}
Here, $\Delta A_\bk$ and $\Delta C_\bk^\text{reg}$ are the differences of the corresponding terms in the evolution of cold and warm particles in numerical simulations.

Since the particles initially are at the same positions in two simulations, initially we have $\Delta \delta_\bk = 0$. The right-hand side of \eqref{exact} can be regarded as a peculiar source for its left-hand side, which gradually generates nonzero $\Delta \delta_\bk$. The term $\Delta \left( A_\bk - C_\bk^\text{reg} \right)$ depends on the exact particle dynamics in the two cases. This term, however, is equal to zero initially, while the total source on the right-hand side is initially given by the remaining two terms. We will see in Sec.~\ref{sec:solution} that the effect of thermal velocities develops on the timescale equal to the initial times $t_\text{in}$ and is caused by the perturbation of the initial velocity $\Delta \dot \delta_\bk$.  This suggests that the effect can be estimated by omitting the last term, $\Delta \left( A_\bk - C_\bk^\text{reg} \right)$, in \eqref{exact} during the whole evolution. One might worry that this assumption may become illegitimate especially during dynamical clustering, where the usual (nonthermal) peculiar velocities of particles may become large, strongly contributing to the term $C_\bk^\text{reg}$. However, the contribution to $A_\bk$ and $C_\bk^\text{reg}$ at small wave numbers $\bk$ from virialized clusters (with the spatial scale $2 \pi / k$ well exceeding the size of a cluster), in fact, by a large part cancel each other \cite{Peebles-AA, Peebles:1980yev}. Therefore, omitting the term $\Delta \left( A_\bk - C_\bk^\text{reg} \right)$ looks reasonable  at relatively small wave numbers for an estimate that follows. For clustering on spatial scales larger or comparable to $2 \pi / k$, what we essentially neglect is the nonlinear contribution to the deviation between the spectra with and without thermal velocities. We note that, for large enough wave numbers, the effects of resolution and shot noise will anyway dominate in the spectrum. The main aim of this paper is to determine the boundary in $k$-space where this occurs.

Other simplifications of equation \eqref{exact} can be made for wave numbers 
\begin{equation}\label{condition}
k \ll k_\text{th} \, ,
\end{equation}
which are typically of interest in simulations.  In this case, using \eqref{kth}, we have
\begin{equation}
4 \pi G \rho = \left( \frac{k_\text{th} v_\text{th}}{a} \right)^2 \gg \left( \frac{k v_\text{th}}{a} \right)^2 \, .
\end{equation}
Thus, the second term in the coefficient of $\Delta \delta_\bk$ on the left-hand side of \eqref{exact} can be neglected. 

Compare now the second term on the left-hand side of \eqref{exact},
\begin{equation}\label{dotterm}
2 H \Delta \dot \delta_\bk = - 2 \ri H \sum_i \frac{m_i}{M} \bk \dot \bx_i^\text{th} e^{\ri \bk \bx_i} \, , 
\end{equation}
with the last term on its right-hand side. The random velocity-dependent factors under the sum in both terms have zero average, but the dispersion of the factor in \eqref{dotterm} is much larger than that of the corresponding factor in the last term of \eqref{exact}. Indeed, using \eqref{kth}, we obtain
\begin{equation}
\overline{\left( 2 H \bk \dot \bx_i^\text{th} \right)^2} = 4 H^2 \left( \frac{k v_\text{th}}{a} \right)^2 = \frac83 \left( \frac{k_\text{th} v_\text{th}}{a} \right)^2 \left( \frac{k v_\text{th}}{a} \right)^2 \, ,
\end{equation} 
while the dispersion of $\xi_i$ is given by \eqref{xidis}.  Hence, the last term in Eq.~\eqref{exact} can also be neglected under condition \eqref{condition}.

Finally, with all these approximations, Eq.~\eqref{exact} takes the simple form
\begin{equation} \label{eff-noise}
\Delta \ddot \delta_\bk + 2 H \Delta \dot \delta_\bk - 4 \pi G \varrho \Delta \delta_\bk =  \left( \frac{k v_\text{th}}{a} \right)^2 \delta_\bk^\text{cold} \, .
\end{equation}
This is our main equation to be solved.

\section{Effects of thermal velocities} \label{sec:noise}

\subsection{Initial conditions}

Let us discuss the initial conditions for Eq.~\eqref{eff-noise}. As described above, the initial positions of particles in the profiles of $\delta^\text{cold}$ and $\delta^\text{warm}$ are the same; hence, their initial values are also the same, giving the initial condition $\Delta \delta_\bk (t_\text{in}) = \delta_\bk^\text{cold} (t_\text{in}) - \delta_\bk^\text{warm} (t_\text{in}) = 0$. What distinguishes the initial conditions is the presence of thermal velocities of particles in the profile $\delta^\text{warm}$. This leads to a nonzero initial value $\Delta \dot \delta_\bk (t_\text{in})$. We have
\begin{equation} \label{noise}
\Delta \dot \delta_\bk  = \dot \delta_\bk^\text{cold} - \dot \delta_\bk^\text{warm} = - \ri \sum_i \frac{m_i}{M} \bk \dot \bx_i^\text{th} e^{\ri \bk \bx_i} \, ,
\end{equation}
with randomly distributed thermal components  $\dot \bx_i^\text{th}$ of particle velocities. Because the thermal velocities $\dot \bx_i^\text{th}$ with different $i$ are independent, the dispersion of quantity \eqref{noise} over realizations of thermal velocities at the initial moment of time is
\begin{equation} \label{growdis}
\overline{ \left| \Delta \dot \delta_\bk (t_\text{in}) \right|^2 } = \frac{1}{N} \left( \frac{k v_\text{in}}{a_\text{in}} \right)^2 \, ,
\end{equation}
where $v_\text{in}$ is the initial value of $v_\text{th}$, and $a_\text{in}$ is the initial value of the scale factor. We thus conclude that, as thermal initial velocities are added to a system with finite number of particles, their main effect is connected with the perturbation of the initial value of the first derivative $\dot \delta_\bk$.

\subsection{Solution}
\label{sec:solution}

For the thermal part of peculiar velocity, we can use the free-streaming approximation. This approximation can be justified by the condition $a H \left| \dot \bv_i^\text{th} \right| \gg \left| \nabla \varphi \left( \bx_i \right)  \right|$, typically valid for thermal velocities. Then, given the equation of motion \eqref{eqmot}, we have $\ddot \bx_i^\text{th} + 2 H \dot \bx_i^\text{th} \simeq 0$, which entails $\dot \bx_i^\text{th} \propto a^{-2}$ and $\bv_i^\text{th} \equiv a \dot \bx_i^\text{th} \propto a^{-1}$.

Using then the laws $v_\text{th} \propto a^{-1}$ and $\delta_\bk^\text{cold} \propto a \propto t^{2/3}$ at the matter-dominated stage, the solution of Eq.~\eqref{eff-noise} with the initial conditions $\Delta \delta_\bk \left( t_\text{in} \right) = 0$ and $\Delta \dot \delta_\bk (t_\text{in}) \ne 0$ is given by
\begin{equation} \label{deltagr}
\Delta \delta_\bk = \left[\frac{9}{10} \left( \frac{t}{t_\text{in}} \right)^{2/3} + \frac35 \left( \frac{t_\text{in}}{t} \right) - \frac32 \right] \left( \frac{k v_\text{in} t_\text{in}}{a_\text{in}} \right)^2 \delta_\bk^\text{cold} ( t_\text{in} ) + \frac35 t_\text{in} \Delta \dot \delta_\bk (t_\text{in}) \left[ \left( \frac{t}{t_\text{in}} \right)^{2/3} - \frac{t_\text{in}}{t} \right] \, .
\end{equation}
At $t \gg t_\text{in}$, the leading contributions are given by the first terms in each of the brackets, proportional to the growing mode of the CDM solution. Since the right-hand side of \eqref{eff-noise} rapidly decays with time, the solution asymptotically continues as a growing mode of the general solution of \eqref{eff-noise} with zero right-hand side.  In this regime, $\Delta \delta_\bk$ respects the same equation as $\delta_\bk^\text{cold}$, so that their solutions on the subsequent stage, where the cosmological constant takes over, will be proportional to each other.  This enables us to write, for $t \gg t_\text{in}$, 
\begin{equation} \label{result-noise}
\Delta \delta_\bk \left( t \right) \simeq \frac{1}{5} \left[ \frac{3 k^2}{k_\text{th}^2 (t_\text{in})} \delta_\bk^\text{cold} \left( t_\text{in} \right) + \frac{2 \Delta \dot \delta_\bk (t_\text{in})}{H_\text{in} } \right] \frac{D (t)}{D (t_\text{in}) } \, , \quad t \gg t_\text{in} \, ,
\end{equation}
where $D (t)$ is the growth factor in the evolution of dark-matter perturbations in $\Lambda$CDM cosmology.

The second term in the bracket of \eqref{result-noise} is the contribution from the noise connected with the finiteness of the $N$-body system.  This is a random quantity with dispersion determined by \eqref{growdis}:
\begin{equation} \label{veldis}
\left\langle \left| \frac{\Delta \dot \delta_\bk (t_\text{in})}{H_\text{in} } \right|^2 \right\rangle = \frac{1}{N} \left( \frac{k v_\text{in}}{a_\text{in} H_\text{in}} \right)^2 = \frac{3}{2 N} \frac{k^2}{k_\text{th}^2 (t_\text{in})} \, .
\end{equation}
Being specified at the initial moment of time $t_\text{in}$, at which the initial particle thermal velocities in \eqref{noise} are uncorrelated with their initial positions, the quantities $\delta_\bk^\text{cold} \left( t_\text{in} \right)$ and $\Delta \dot \delta_\bk (t_\text{in})$ are also statistically uncorrelated.

Defining the difference between the corresponding power spectra as
\begin{equation}
\Delta P (k, z) = P_\text{warm} (k, z) - P_\text{cold} (k, z) \, ,
\end{equation}
and statistically averaging it over realizations of initial conditions [which amounts to averaging over realizations of $\dot \delta_\bk (t_\text{in})$ using \eqref{veldis}], we will have, for $1 + z \ll 1 + z_\text{in}$,
\begin{align} \label{wdm-spec}
\Delta P (k, z)  &\simeq \left[ - \frac{6 k^2}{5 k_\text{th}^2 (z_\text{in})} P_\text{cold} (k, z_\text{in}) + \frac{6 V}{25 N} \frac{k^2}{k_\text{th}^2 (z_\text{in})} \right] \left[ \frac{D (z)}{D (z_\text{in})} \right]^2 \nonumber \\ 
&= - \frac{6 k^2}{5 k_\text{th}^2 (z_\text{in})} P_\text{cold} (k, z) + \frac{6 V}{25 N} \frac{k^2}{k_\text{th}^2 (z_\text{in})} \left[ \frac{D (z)}{D (z_\text{in})} \right]^2 \,,
\end{align}
where $D (z)$ is the growth factor as a function of redshift $z$. Thus, for $1 + z \ll 1 + z_\text{in}$, we have
\begin{equation}\label{main1}
P_\text{warm} (k, z) = \left[ 1 - \frac{6 k^2}{5 k_\text{th}^2 (z_\text{in})} \right] P_\text{cold} (k, z) + \frac{6}{25} \frac{V}{N} \frac{k^2}{k_\text{th}^2 (z_\text{in})} \left[ \frac{D (z)}{D (z_\text{in})} \right]^2 \, .
\end{equation}

The exact solution for $D (z)$ in the case of universe filled with dark matter and cosmological constant can be found in \cite[Sec.~6.3.4]{Mukhanov:2005sc}. It can be very well approximated by the simple form
\begin{equation} \label{growing}
D (z) = \frac{1}{1 + z} \left[ \Omega_\text{m} + \frac{\Omega_\Lambda}{(1 + z)^3} \right]^{-1/3} \, , 
\end{equation}
which is normalized as $D (0) = 1$.  Here, $\Omega_\text{m}$ and $\Omega_\Lambda$ are the usual cosmological parameters for matter and cosmological constant, respectively.  We will use this form of the growing factor in our plots below.

If the interval between $z_\text{in}$ and $z$ is not that large, then we can turn to solution \eqref{deltagr} with all terms present to get corrected estimates for the effects of thermal velocities under consideration. The physical effect (the one that survives in the limit $N \to \infty$) is then given by
\begin{equation} \label{phys}
\Delta_\text{phys} P (k, z)  \simeq - \frac{6 k^2}{5 k_\text{th}^2 (z_\text{in})} P_\text{cold} (k, z) \left[ 1 + \frac23 \left( \frac{1 + z}{1 + z_\text{in}} \right)^{5/2} - \frac53 \left( \frac{1 + z}{1 + z_\text{in}} \right) \right] \, ,
\end{equation}
and the dominating contribution connected with the perturbation of the initial time derivative $\dot \delta_\bk$ is
\begin{equation} \label{art}
\Delta_N P ( k, z ) \simeq \frac{6 V}{25 N} \, \frac{k^2}{k_\text{th}^2 (z_\text{in})} \left[ \frac{D (z)}{D (z_\text{in})} \right]^2  \left[1 - \left( \frac{1 + z}{1 + z_\text{in}} \right)^{5/2} \right]^2 \, .
\end{equation}
Then
\begin{equation}\label{main2}
P_\text{warm} (k, z) = P_\text{cold} (k, z) + \Delta_\text{phys} P (k, z) + \Delta_N P ( k, z ) \, .
\end{equation}

Equations \eqref{main1} and \eqref{main2} are our main results.

The power spectrum $P_\text{warm} (k, z)$ [as well as $P_\text{cold} (k, z)$] in $N$-body simulations also contains the shot-noise contribution \eqref{shot}, which has to be taken into account.

\subsection{Physical effect}
\label{sec:regeff}

The first, $N$-independent, terms on the right-hand side of \eqref{main1} or \eqref{main2} describe the physical effect of thermal velocities that survives in the limit of $N \to \infty$. 

For dark matter in the form of thermal relic particles with the distribution function \eqref{fermi} with $\chi = 1$, the thermal velocity, according to definition \eqref{veldisp}, is given by \cite{Leo:2017zff} (in units of the speed of light)
\begin{equation}\label{vinth}
v_\text{th} (z) = 0.8 \times 10^{-9} \left( \frac{\omega_\text{DM}}{0.12} \right)^{1/3}	\left( \frac{\text{keV}}{m_\text{TR}} \right)^{4/3} (1 + z) \, ,
\end{equation}
where $\omega_\text{DM} = \Omega_\text{DM} h^2$, and this quantity is normalized by the current best-fit value $\omega_\text{DM} \approx 0.12$. For dark matter composed of thermal sterile neutrino, the thermal velocity, calculated according to \eqref{veldisp}, is
\begin{equation} \label{vin}
v_\text{th} (z) = \sqrt{\frac{5 \zeta(5)}{\zeta(3)}} \frac{T_\nu (z)}{m_\text{SN}} \approx \frac{2.077 \, (4/11)^{1/3} T_\gamma (1 + z)}{m_\text{SN}} = 3.5 \times 10^{-7}\, (1 + z)\, \frac{\text{keV}}{m_\text{SN}} \, .
\end{equation}
This can be expressed through the redshift of nonrelativistic transition $z_\text{nr}$ defined via the condition
\begin{equation} \label{znr}
\sqrt{3} v_\text{th} (z) = \frac{1 + z}{1 + z_\text{nr}} \, , \qquad 1 + z \ll 1 + z_\text{nr} \, .
\end{equation}
Then, according to \eqref{vinth} and \eqref{vin}, we have
\begin{equation}
1 + z_\text{nr} \simeq 0.7 \times 10^9\, \left( \frac{0.12}{\omega_\text{DM}} \right)^{1/3} \left( \frac{m_\text{TR}}{\text{keV}} \right)^{4/3} \, , \qquad 1 + z_\text{nr} \simeq 1.7 \times 10^6\, \frac{ m_\text{SN}}{\text{keV}} \, , 
\end{equation}
respectively, for thermal relic and sterile neutrino, which determines the redshift values below which our theory of nonrelativistic particles will be applicable.  Since numerical simulations with warm dark matter typically are performed at $z \lesssim 200$, these conditions will be satisfied for all reasonable masses of dark-matter particles.

The thermal wave number \eqref{kth} is calculated to be
\begin{align} \label{kth-th}
k_\text{th} (z) &\approx \frac{1.7 \times 10^3}{\sqrt{1 + z}} \left( \frac{\omega_\text{DM}}{0.12} \right)^{1/6} \left( \frac{m_\text{TR}}{\text{keV}} \right)^{4/3} \, \text{Mpc}^{-1} \, , \\
k_\text{th} (z) &\approx \frac{4 \times 10^2}{\sqrt{1 + z}} \left( \frac{\omega_\text{DM}}{0.12} \right)^{1/2} \left( \frac{m_\text{SN}}{\text{keV}} \right) \, \text{Mpc}^{-1} \, , \label{kth-sn}
\end{align}
respectively. The physical effect on the power spectrum for $1 + z \ll 1 + z_\text{in}$ is, respectively,
\begin{align} \label{regeff}
\frac{\Delta_\text{phys} P \left( k, z \right)}{P_\text{cold} \left( k, z \right)} &\simeq - 3 \times 10^{-11} \left( \frac{0.12}{\omega_\text{DM}} \right)^{1/3} \left( \frac{k}{\text{Mpc}^{-1}} \right)^2 \left( \frac{\text{keV}}{m_\text{TR}} \right)^{8/3} (1 + z_\text{in}) \, , \\	 
\frac{\Delta_\text{phys} P \left( k, z \right)}{P_\text{cold} \left( k, z \right)} &\simeq - 6 \times 10^{-6} \left( \frac{0.12}{\omega_\text{DM}} \right) \left( \frac{k}{\text{Mpc}^{-1}} \right)^2 \left( \frac{\text{keV}}{m_\text{SN}} \right)^2 (1 + z_\text{in}) \, . 
\end{align}
Setting in the last equation $z_\text{in} = 199$, $k = 20\,\text{Mpc}^{-1}$, and $m_\text{SN} = 2\, \text{keV}$, we get relative correction to the power spectrum at the level of $10^{-1}$, the smallness of which justifies condition \eqref{condition} in view of \eqref{phys}.

\subsection{Effects of discreteness}
\label{sec:resol}

For finite $N$, adding thermal velocities to the simulation particle results in (unwanted) numerical effects at relatively large $k$, described by the last terms in \eqref{main1} and \eqref{main2}. In this section, we plot several graphs demonstrating these effects in the case of sterile neutrino. 

\begin{figure}[htp]
\begin{center}
\includegraphics[width=.66\textwidth]{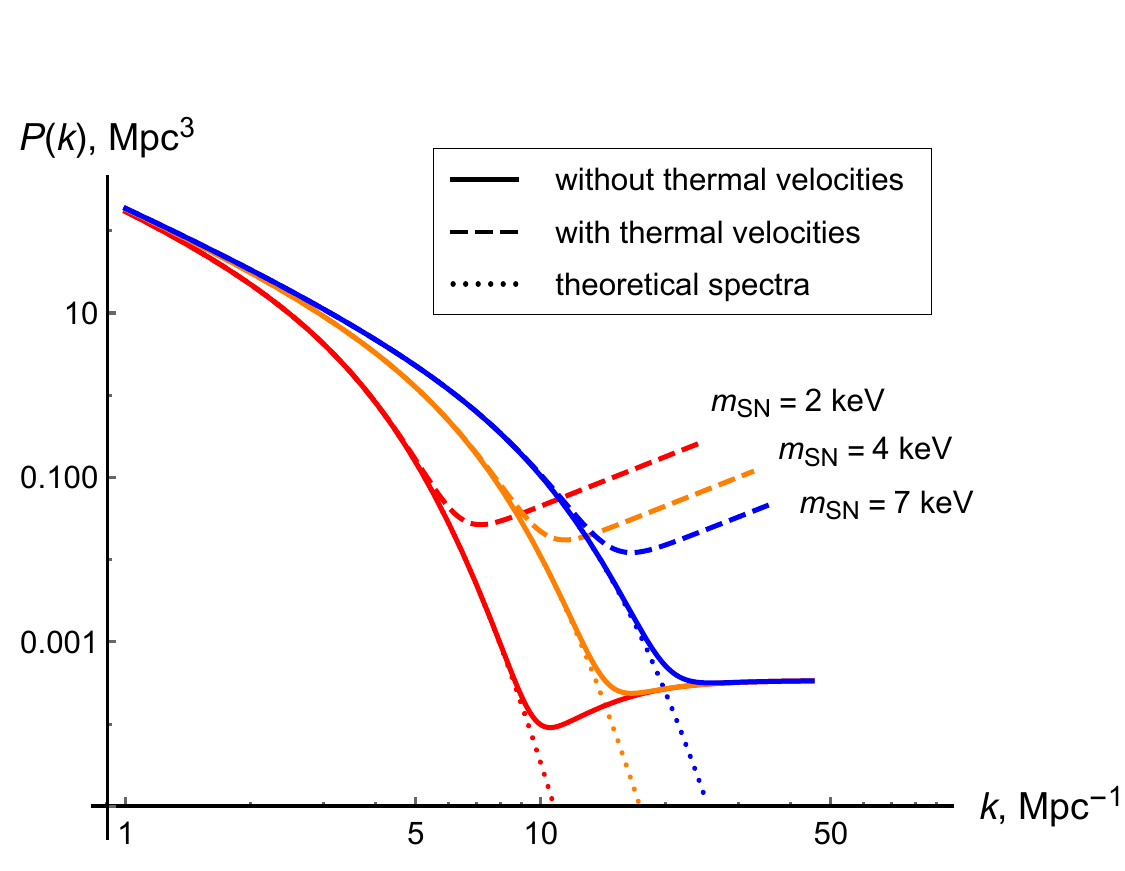}
\end{center}
\caption{Plots of the resulting analytic power spectrum \eqref{main1} at $z = 0$ shown for sterile neutrino of different masses $m_\text{SN} = 2$, $4$, and $7$~keV. Solid curves correspond to spectra without thermal velocities (but with the shot-noise contribution \eqref{shot} included), and dashed curves show the effect of switching thermal velocities at the initial redshift $z_\text{in}$. Dotted continuations are the theoretical spectra \eqref{wdmpower}. All curves are plotted for $\omega_\text{DM} = 0.12$, $h = 0.7$, the initial redshift $z_\text{in} = 199$, the box size $L = 25\, h^{-1}\,\text{Mpc}$ and $N = 512^3$. \label{fig:SN_mass}}
\end{figure}

\begin{figure}[htp]
\begin{center}
\includegraphics[width=.66\textwidth]{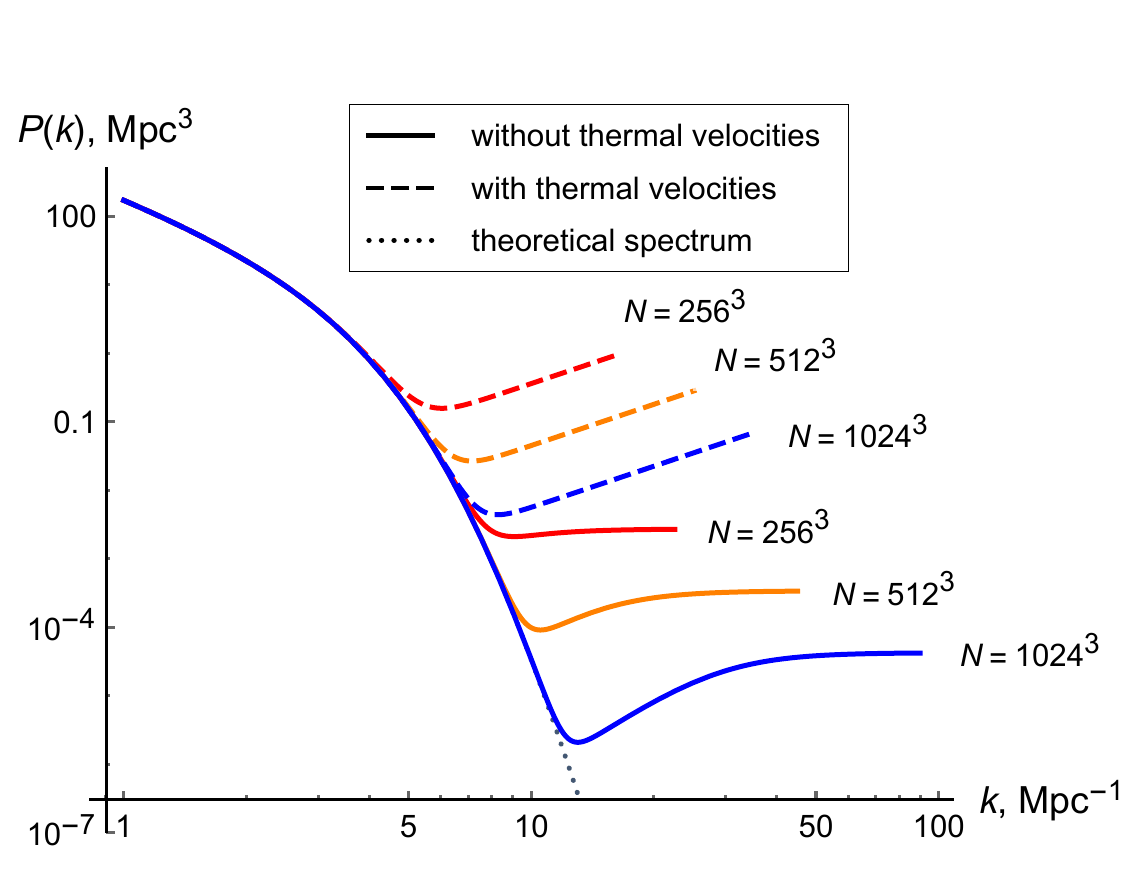}
\end{center}
\caption{Plots of the resulting analytic power spectrum \eqref{main1} at $z = 0$ shown for sterile neutrino of fixed mass and for different numbers of simulation particles $N = 256^3$, $512^3$, and $1024^3$. Solid curves correspond to spectrum without thermal velocities (but with the shot-noise contribution \eqref{shot} included), and dashed curves show the effect of switching thermal velocities at the initial redshift $z_\text{in}$. Dotted continuation is the theoretical spectrum \eqref{wdmpower}. All curves are plotted for $\omega_\text{DM} = 0.12$, $h = 0.7$, the initial redshift $z_\text{in} = 199$, the box size $L = 25\, h^{-1}\,\text{Mpc}$ and the sterile-neutrino mass $m_\text{SN} = 2\,\text{keV}$. \label{fig:SN_N}}
\end{figure}

\begin{figure}[htp]
\begin{center}
\includegraphics[width=.66\textwidth]{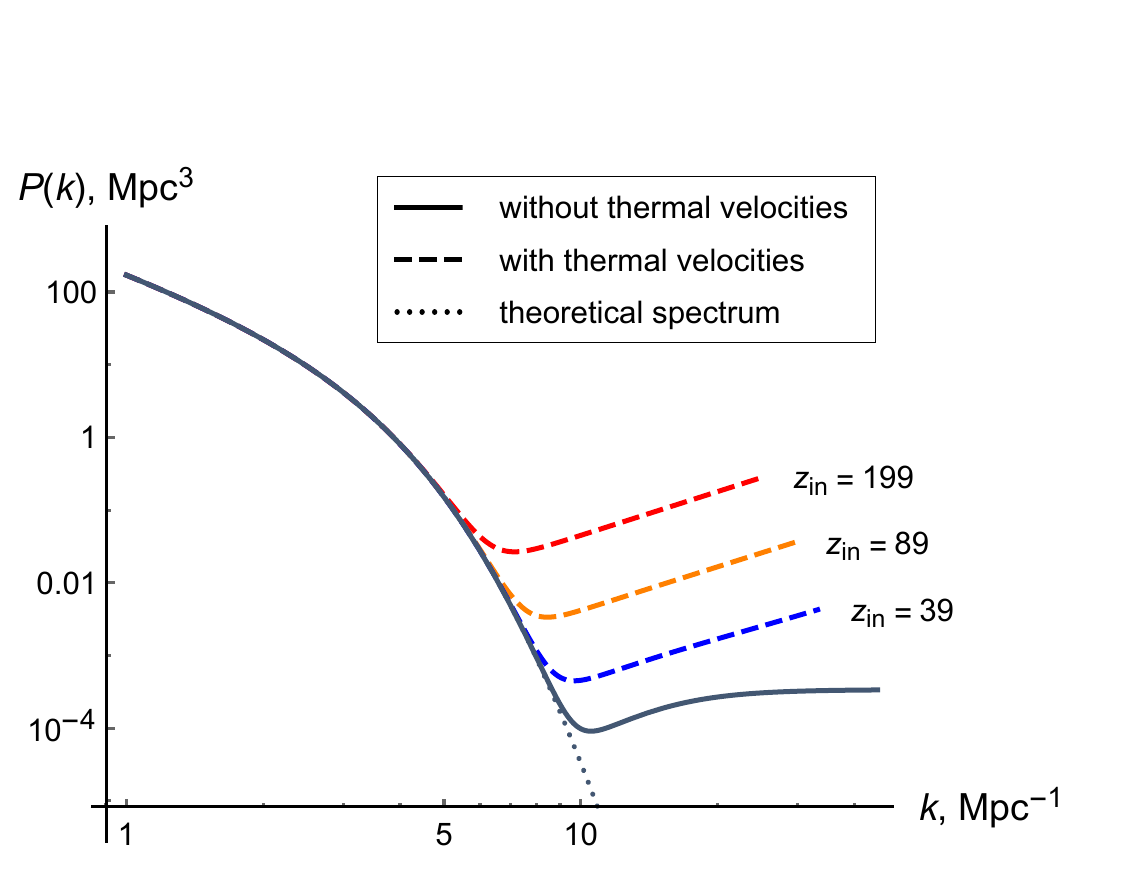}
\end{center}
\caption{Plots of the resulting analytic power spectrum \eqref{main1} at $z = 0$ shown for different values of $z_\text{in} = 39$, $89$, and $199$, at which the thermal velocities are switched on. Solid curve corresponds to spectrum without thermal velocities (but with the shot-noise contribution \eqref{shot} included), and dashed curves show the effect of switching thermal velocities. Dotted continuation is the theoretical spectrum \eqref{wdmpower}. All curves are plotted for $\omega_\text{DM} = 0.12$, $h = 0.7$, the box size $L = 25\, h^{-1}\,\text{Mpc}$, the particle number $N = 512^3$, and the sterile-neutrino mass $m_\text{SN} = 2\,\text{keV}$. \label{fig:SN_z}}
\end{figure}

In Figs.~\ref{fig:SN_mass}--\ref{fig:SN_z}, solid curves show power spectra $P(k)$ without the thermal velocities (but with the shot-noise contribution \eqref{shot} included), which are plotted up to the Nyquist wave number \eqref{kN} for corresponding values of the box size $L = V^{1/3}$ and particle number $N$. Dashed curves in these figures show power spectra with thermal velocities and are plotted till $k = k_\text{th} \left( z_\text{in} \right) k_N /\left[ k_\text{th} \left( z_\text{in} \right) + k_N \right]$, so that $k$ does not exceed either $k_\text{th} \left( z_\text{in} \right)$ or $k_N$, and our approximation \eqref{condition} remains valid. Dotted continuations are the theoretical linear spectra \eqref{wdmpower} for sterile neutrino. 

Figure~\ref{fig:SN_mass} shows the effect of sterile-neutrino mass; here three different plots correspond to different neutrino masses, and the parameters are indicated in the figure caption.  Solid curves describe the expected resulting power spectrum in simulations without adding initial thermal velocities.  Their deviation from the theoretical power spectrum at higher values of $k$ is caused by the shot noise \eqref{shot}.  Dashed curves describe the linearly extrapolated resulting power spectrum as thermal velocities are switched on at redshift $z_\text{in}$. One observes a turnover in the power spectrum at certain wave number, after which it starts growing as $P_\text{warm} (k) \propto k^2$, corresponding to the last term in \eqref{main1}.  The wave number $k_*$ of turnover can be estimated by comparing this term to the first term on the right-hand side of \eqref{main1}, which amounts to solving the equality
\begin{equation}\label{k*}
P_\text{WDM}  \left( k_*, z_\text{in} \right) = \frac{6}{25} \frac{V}{N} \frac{k_*^2}{k_\text{th}^2 (z_\text{in})} \, .
\end{equation}
The expected results of simulations at $k \gtrsim k_*$ contain numerical artifacts. 

The characteristic behavior $P_\text{warm} (k) \propto k^2$ in the power spectra at $k \gtrsim k_*$ are observed in real numerical simulations (e.g., in \cite{Colin:2007bk, Leo:2017zff}).

Similarly, Figure~\ref{fig:SN_N} shows the expected effect of thermal velocities for different values of $N$, and Figure~\ref{fig:SN_z} for different values of $z_\text{in}$.

\begin{figure}[htp]
\begin{center}
\includegraphics[width=.66\textwidth]{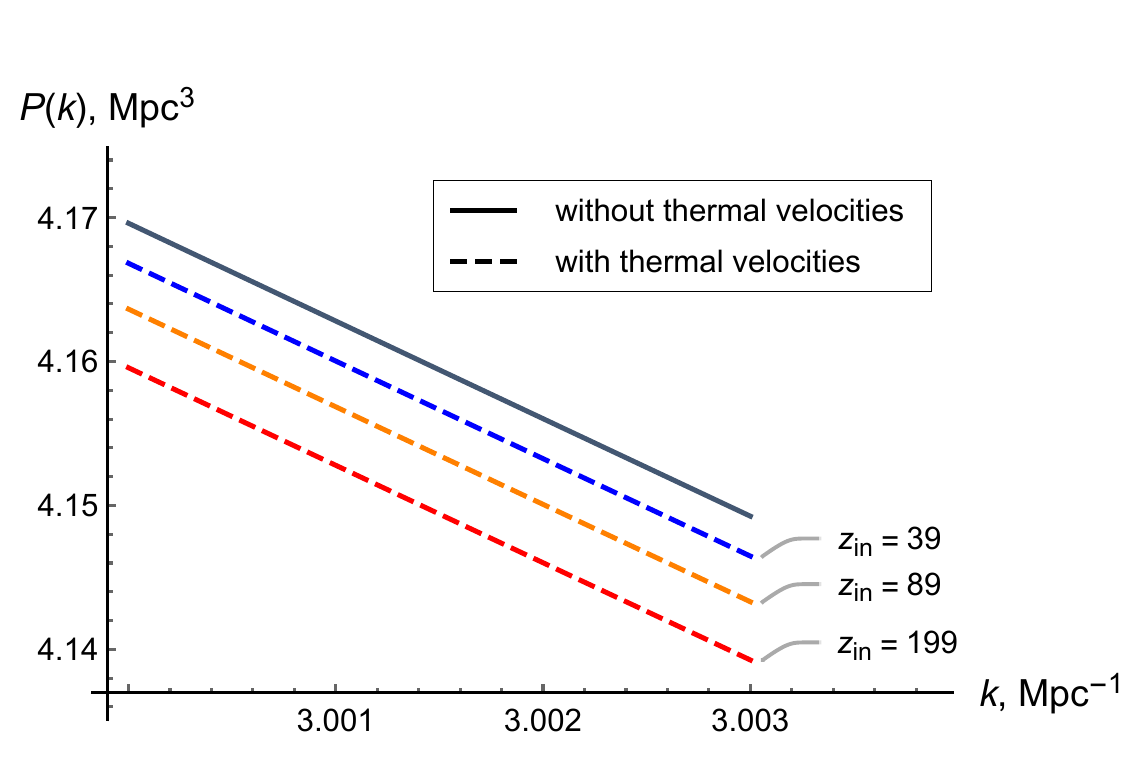}
\end{center}
\caption{Zoom of a small region of the plot in Fig.~\ref{fig:SN_z}. Solid curve corresponds to the spectrum without thermal velocities, and dashed curves show the result of switching thermal velocities at different initial redshifts $z_\text{in} = 39$, $89$, and $199$. This figure captures the physical effect \eqref{phys} of suppression of the power spectrum in the region of $k$ well below the turnover scale $k_*$, where the effects of resolution are subdominant. The power spectrum becomes the more suppressed the earlier one turns on thermal velocities. \label{fig:SN_zoom}}
\end{figure}

The small physical effect of thermal velocities at $k \lesssim k_*$, that suppresses power spectrum on these scales, is demonstrated in Fig.~\ref{fig:SN_zoom}, which shows a small region of scales for the plot of Fig.~\ref{fig:SN_z}. The earlier the thermal velocities are turned on, the more prominent is their effect in simulations.

As we have seen, the main disturbance of the spectrum caused by thermal velocities comes from the uncontrolled noise in the initial condition \eqref{noise}.  One might try do reduce this noise by adding thermal velocities not randomly but with correlations among neighbors in such a way that they average to zero over certain small clusters of simulation particles.  For instance, after choosing randomly a thermal velocity for the first particle, one could assign exactly opposite thermal velocity to its randomly chosen neighbor, and then repeat this procedure with other particle pairs. One could also do the same with triples of particles, and so on. Whether such scheme is going to work requires special investigation with simulations.

\subsection{Comparison with numerical simulations}

\begin{figure}[htp]
\centering
\subfigure[][$N=512^3$, $m_\mathrm{WDM} = 3.3$~keV. Varying $L$.]{\includegraphics[width=.49\textwidth]{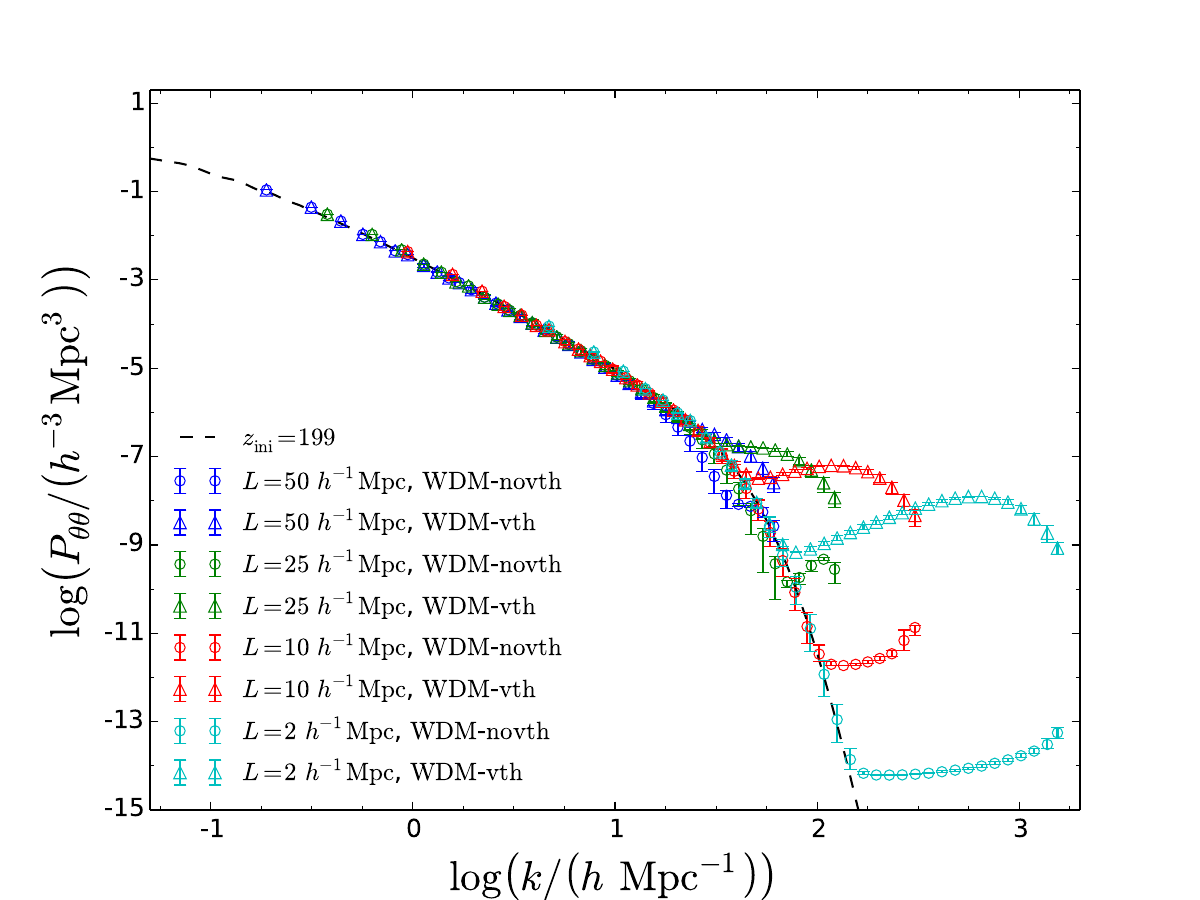}\label{fig:a}} 
\subfigure[][$L = 2\, h^{-1}$\,Mpc, $m_\mathrm{WDM} = 3.3$~keV. Varying $N$.] {\includegraphics[width=.49\textwidth]{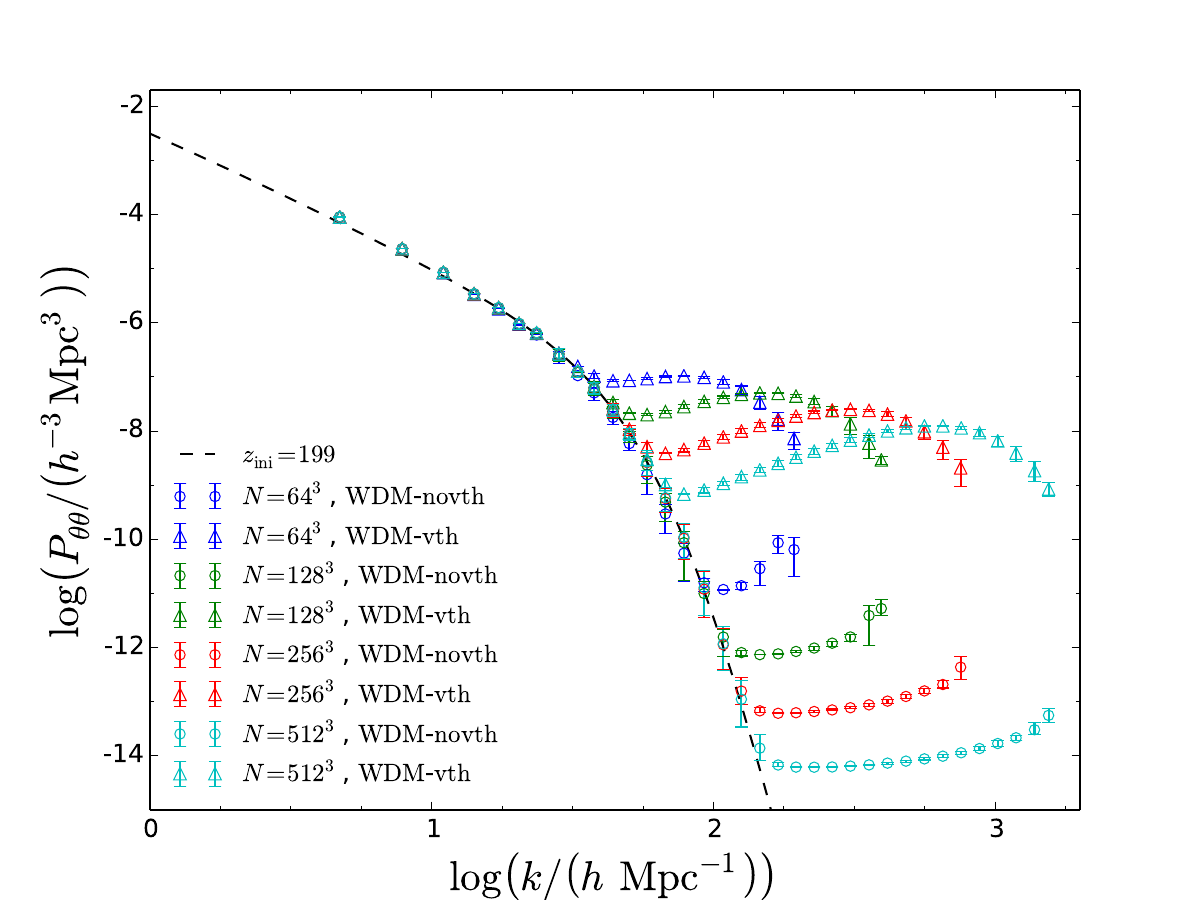}\label{fig:b}} 
\subfigure[][$N=512^3$, $L = 2\, h^{-1}$\,Mpc. Varying $m_\mathrm{WDM}$.]{\includegraphics[width=.49\textwidth]{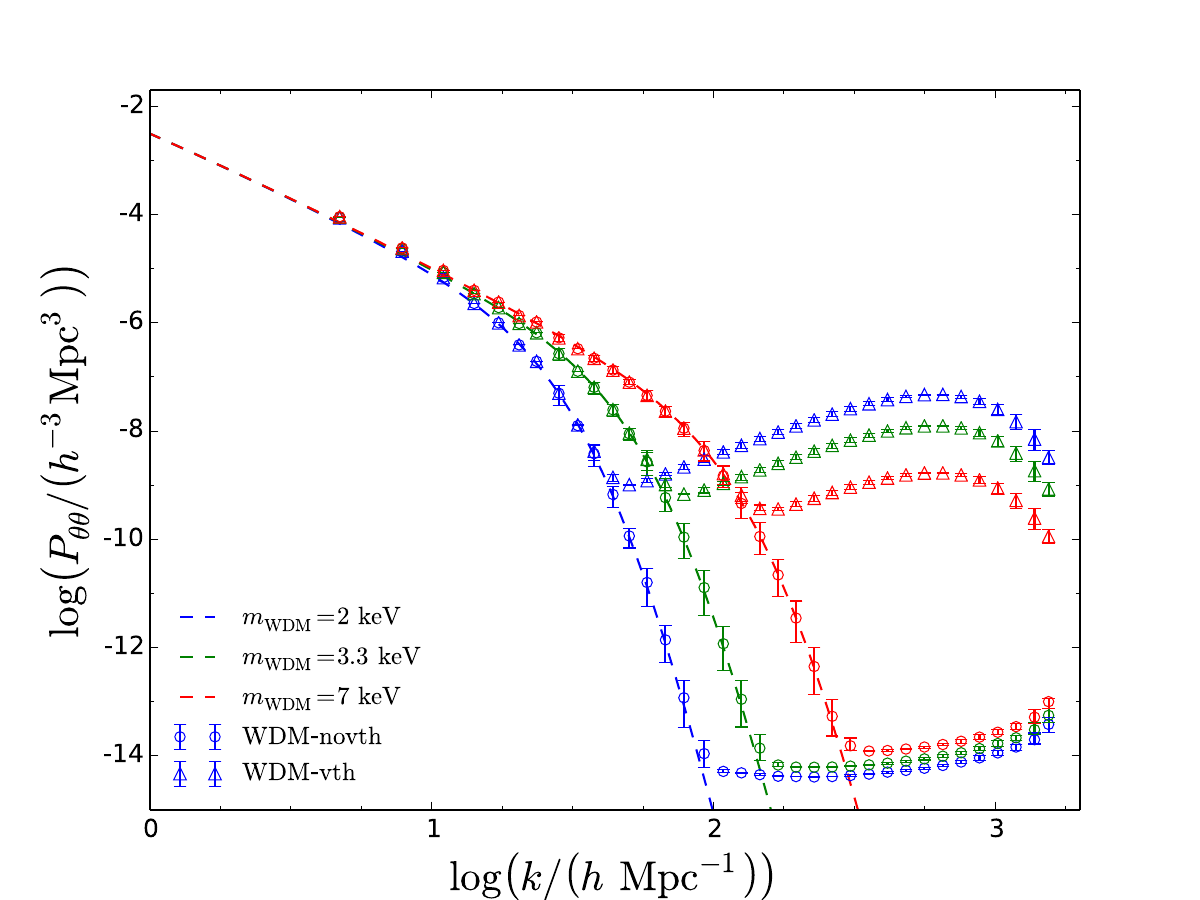}\label{fig:c}} 
\caption{Power spectra $P_{\theta\theta}$ measured in the simulations of \cite{Leo:2017zff} from initial conditions for WDM with thermal velocities (WDM-vth, triangles) and without thermal velocities (WDM-novth, dots). The different panels show how the spectra change when varying (a)~the box length $L$, (b)~the number of particles $N$ and (c)~the mass of thermal relic $m_\text{TR} = m_\text{WDM}$, while fixing the other parameters. The velocity power spectra are normalized such that $P_{\theta\theta} = P_\text{WDM}$, so that they can be compared with the theoretical linear matter power spectra represented as dashed lines. This is Fig.~2 from \cite{Leo:2017zff}, which is reproduced here with the consent of the authors. \textcopyright~IOP Publishing. Reproduced with permission. All rights reserved.}
\label{fig:simul}
\end{figure}

As a test of our model, we compare some of its predictions with the numerical results presented in Fig.~2 of \cite{Leo:2017zff}, which we reproduce here as Fig.~\ref{fig:simul}. There, the initial conditions with and without thermal velocities were set at $z_\text{in} = 199$ with variable box size $L$, number of particles $N$ or mass of thermal relic $m_\text{TR}$. It should be noted that the results in Fig.~2 of \cite{Leo:2017zff} are presented for the power spectra of the velocity divergence at the initial moment of time (i.e., at $z_\text{in} = 199$). 

In linear theory (valid at high values of $z$), the divergence of the velocity field on sub-Hubble spatial scales is related to the density contrast as
\begin{equation}
\theta \equiv \nabla \cdot \bv = - a \dot \delta \, .
\end{equation}
For the initial conditions with thermal velocities, there is a random thermal contribution to $\dot \delta$ given by \eqref{noise}, with spectral dispersion given by \eqref{growdis} or \eqref{veldis}. On the other hand, the contribution of regular velocities to $\dot \delta$ is given by the usual growth law: $\dot \delta_\text{cold} = H \delta$.  Hence, using \eqref{veldis}, the initial velocity power spectrum can be approximated as
\begin{equation}\label{Ptt}
P_{\theta\theta} \left( k , z_\text{in} \right) = P_\text{WDM} \left( k , z_\text{in} \right)  + \frac{3 V}{2 N} \frac{k^2}{k_\text{th}^2 (z_\text{in})} + P_{N\theta\theta} (k) \, .
\end{equation}
Here, we have normalized the initial power spectrum $P_{\theta\theta} (k)$ in such a way that it coincides with $P_\text{WDM} (k)$ at small $k$, as this is also done in \cite{Leo:2017zff}. We have also included the subdominant Nyquist shot noise $P_{N \theta\theta} (k)$ in \eqref{Ptt}, which will be of no interest to us here. Expression \eqref{Ptt} qualitatively matches the behavior observed in numerical simulations of \cite{Leo:2017zff} presented in Fig.~\ref{fig:simul}, while its features are inherited in the evolved density power spectrum \eqref{main1} and \eqref{main2}.

To determine the turnover scale $k_\star$ at which thermal velocities are going to take over in the initial velocity power spectrum, we should compare the first and second terms on the right-hand side of \eqref{Ptt}. We then obtain an equation similar to \eqref{k*} with a somewhat different coefficient:
\begin{equation}\label{kvel}
P_\text{WDM} \left(k_\star, z_\text{in} \right) = \frac{3 V}{2 N} \frac{k_\star^2}{k_\text{th}^2 (z_\text{in})} \, .
\end{equation}
To make a comparison with the results of \cite{Leo:2017zff}, we calculate the values of $\log_{10} k_\star$ for $k_\star$ in units of $h / \text{Mpc}$ in each situation using Eq.~\eqref{kvel} and the theoretical linear power spectrum evaluated at $z_\text{in}$ by using \eqref{wdmpower} and the growth factor \eqref{growing}.

For simulations with $N = 512^3$, $m_\text{TR} = 3.3$~keV, and variable box size $L = 50\, h^{-1}$~Mpc, $25\, h^{-1}$~Mpc, $10\, h^{-1}$~Mpc, and $2\, h^{-1}$~Mpc, our Eq.~\eqref{kvel} gives, respectively, $\log_{10} k_\star \approx 1.38$, 1.51, 1.66, and 1.84, in a good agreement with the results displayed in Fig.~2(a) of \cite{Leo:2017zff} [reproduced here as Fig.~\ref{fig:a}].

For simulations with $L = 2\, h^{-1}$~Mpc, $m_\text{TR} = 3.3$~keV, and variable particle number $N = 64^3$, $128^3$, $256^3$, and $512^3$, our equation \eqref{kvel} gives, respectively, $\log_{10} k_\star \approx 1.58$, 1.69, 1.77, and 1.84, in a good agreement with the results displayed in Fig.~2(b) of \cite{Leo:2017zff} [reproduced here as Fig.~\ref{fig:b}].

For simulations with $L = 2\, h^{-1}$~Mpc, $N = 512^3$, and variable masses of thermal relics $m_\text{TR} = 2$~keV, 3.3~keV, and 7~keV, our Eq.~\eqref{kvel} gives, respectively, $\log_{10} k_\star \approx 1.64$, 1.84, and 2.13, in a good agreement with the results displayed in Fig.~2(c) of \cite{Leo:2017zff} [reproduced here as Fig.~\ref{fig:c}].

Thus, our model predicts a correct turnover scale and qualitative behavior of the spectrum when compared with the numerical simulations of \cite{Leo:2017zff}.

It should be noted that, in numerical simulations, large initial differences between the power spectra with and without initial thermal velocities become gradually much smaller in the course of evolution on spatial scales that enter essentially nonlinear regime \cite{Leo:2017zff}. This effect is not captured by our theory, which essentially works in the linear approximation. In any case, it remains questionable whether such simulations reproduce adequate physical picture on spatial scales that were dominated by the resolution noise in the course of evolution, i.e., with $k > k_\star$. For nonlinear evolution with clustering, perhaps, one should speak about the mass scales corresponding to $k_\star$, which is the dark-matter mass in the homogeneous universe within a sphere of comoving diameter $d = \pi / k_\star$, and which is equal to
\begin{equation}
M_\star = \frac{\pi^3 \Omega_\text{DM} H_0^2}{16 G k_\star^3} \approx 5 \times 10^{11} \left( \frac{\omega_\text{DM}}{0.12} \right) \left( \frac{\text{Mpc}^{-1}}{k_\star} \right)^3 M_\odot \, .
\end{equation} 
Dark-matter halo mass function in simulations is then questionable for halo masses $M < M_\star$.

\section{Matter consisting of several components}
\label{sec:several}

Using the methods of the preceding section, it is easy to obtain equations describing the evolution of density contrasts of nonrelativistic matter consisting of several components with different thermal velocity dispersions. We label the components by index $n$ and introduce the densities for all components according to \eqref{density}:
\begin{equation} \label{density-a}
\rho_n (\bx) = \sum_{i_n} \frac{m_{i_n}}{a^3} \delta \left( \bx - \bx_{i_n} \right) \, , \qquad \delta^{(n)}_{\bk} = \int \frac{ \rho_n (\bx)}{\varrho_n V} e^{\ri\bk\bx} d^3 \bx = \sum_{i_n} \frac{m_{i_n}}{M_n} e^{\ri\bk\bx_{i_n}} \, .
\end{equation}
Here, $\varrho_n$ and $M_n = \sum_{i_n} m_{i_n}$ are the background density and the total mass, respectively, of the $n^\text{th}$ component. The Newtonian potential is then generalized from \eqref{potential} to
\begin{equation} \label{potential-a}
\varphi (\bx) = - 4 \pi G a^2 \sum_n \varrho_n \sum_{\bk \ne 0} e^{- \ri \bk \bx} \frac{\delta^{(n)}_\bk}{k^2} \, .
\end{equation}

Repeating the similar steps that lead to \eqref{nonlin}, one can obtain the following system of equations for the evolution of the Fourier components of the density contrasts:
\begin{equation} \label{nonlin-a}
\ddot \delta^{(n)}_\bk + 2 H \dot \delta^{(n)}_\bk = 4 \pi G \varrho \delta_\bk + A^{(n)}_\bk - C^{(n)}_\bk \, ,
\end{equation}
where
\begin{equation}
\varrho = \sum_n \varrho_n \, , \qquad \varrho \delta_\bk = \sum_n \varrho_n \delta^{(n)}_\bk
\end{equation}
define the total background density and its perturbation, respectively, and
\begin{equation} \label{A-ab}
A^{(n)}_\bk = 4 \pi G \varrho \sum_{\bk' \ne 0, \bk} \frac{\bk \bk'}{k'^2} \delta^{(n)}_{\bk - \bk'} \delta_{\bk'} \, ,
\end{equation}
\begin{equation} \label{C-a}
C^{(n)}_\bk = \sum_{i_n} \frac{ m_{i_n}}{M_n} \left( \bk \dot \bx_{i_n} \right)^2 e^{\ri \bk \bx_{i_n}} \, .
\end{equation}

Since all components of matter are nonrelativistic, we can write
\begin{equation}
\varrho_n = f_n \varrho \, , \qquad \delta_\bk = \sum_n f_n \delta^{(n)}_\bk \, , \qquad \sum_n f_n = 1 \, , \qquad f_n = \frac{M_n}{M} = \text{const} \, .
\end{equation}
For the total density contrast, we have the following equation:
\begin{equation} \label{nonlin-tot}
\ddot \delta_\bk + 2 H \dot \delta_\bk = 4 \pi G \varrho \delta_\bk + A_\bk - \sum_n f_n C^{(n)}_\bk \, ,
\end{equation}
where $A_\bk$ is given by \eqref{A}. The only difference of this equation from
\eqref{nonlin} is the sum over different components in the velocity term on the right-hand side.  

We note that the thermal part of the partial velocity term \eqref{C-a} is similar to \eqref{cregth}, \eqref{cth}:
\begin{equation} \label{caregth}
\left[ C_\bk^{(n)} \right]_\text{th} = \sum_{i_n} \frac{m_{i_n}}{M_n} \left( \bk \dot \bx_{i_n}^\text{th} \right)^2 e^{\ri \bk \bx_{i_n}} = \left( \frac{k v^{(n)}_\text{th}}{a} \right)^2 \delta^{(n)}_\bk + \sum_{i_n} \frac{ m_{i_n}}{M_n}\, \xi_{i_n} e^{\ri \bk \bx_{i_n}} \, ,
\end{equation}
with expression for $\xi_{i_n}$ given in \eqref{xi}. 
Writing the last term in \eqref{nonlin-tot} as the sum of ``regular'' and ``thermal'' contributions, and proceeding along the same lines as in Sec.~\ref{sec:model}, we will obtain a generalization of the effective equation \eqref{eff-noise}:
\begin{equation} \label{eff-noise-n}
\Delta \ddot \delta^{(n)}_\bk + 2 H \Delta \dot \delta^{(n)}_\bk = 4 \pi G \varrho \Delta \delta_\bk + \left( \frac{k v^{(n)}_\text{th}}{a} \right)^2 \delta_\bk^{(n)\,\text{cold}} \, ,
\end{equation}
where $\delta_\bk^{(n)\,\text{cold}}$ is the density profile in the case where all particles are cold, i.e., initial thermal velocities are not added.  The perturbation of the initial condition is
\begin{equation} \label{anoise} 
\Delta \dot \delta^{(n)}_\bk  = \dot \delta_\bk^{(n)\,\text{cold}} - \dot \delta_\bk^{(n)\,\text{warm}} = - \ri \sum_{i_n} \frac{m_{i_n}}{M_n} \bk \dot \bx_{i_n}^\text{th} e^{\ri \bk \bx_i} \, .
\end{equation}
Its initial dispersion is given, similarly to \eqref{growdis}, by
\begin{equation} \label{growdis-n}
\overline{ \left| \Delta \dot \delta^{(n)}_\bk (t_\text{in}) \right|^2 } = \frac{1}{N_n} \left( \frac{k v^{(n)}_\text{in}}{a_\text{in}} \right)^2 \, ,
\end{equation}
where $N_n$ is the number of particles in the $n$th component. Equation for the total density contrast is obtained from \eqref{eff-noise-n} by summing over $n$ with weights $f_n$:
\begin{equation} \label{eff-noise-tot}
\Delta \ddot \delta_\bk + 2 H \Delta \dot \delta_\bk = 4 \pi G \varrho \Delta \delta_\bk + \sum_n f_n \left( \frac{k v^{(n)}_\text{th}}{a} \right)^2 \delta_\bk^{(n)\,\text{cold}} \, .
\end{equation}
Its growth rate is given by
\begin{equation}
\Delta \dot \delta_\bk = \Delta \dot \delta^\text{cold}_\bk - \Delta \dot \delta^\text{warm}_\bk = \sum_n f_n \Delta \dot \delta^{(n)}_\bk \, ,
\end{equation}
with the initial dispersion
\begin{equation} \label{growdis-total}
\overline{ \left| \Delta \dot \delta_\bk (t_\text{in}) \right|^2 } = \sum_n f_n^2 \overline{ \left| \Delta \dot \delta^{(n)}_\bk (t_\text{in}) \right|^2 } = \sum_n \frac{f_n^2}{N_n} \left( \frac{k v^{(n)}_\text{in}}{a_\text{in}} \right)^2 \, ,
\end{equation}

In the simplest case where particles of all components are initially distributed with the same profile, so that $\delta^{(n)\,\text{cold}}_\bk \simeq \delta_\bk^{\text{cold}}$, we obtain a generalization of \eqref{phys} and \eqref{art} for the total power spectrum:
\begin{align} \label{aphys}
\Delta_\text{phys} P (k, z)  &\simeq - \sum_n f_n \frac{6 k^2}{5 k_{\text{th} (n)}^2 (z_\text{in})} P (k, z_\text{in}) \left[ \frac{D (z)}{D (z_\text{in})} \right]^2 \left[ 1 + \frac23 \left( \frac{1 + z}{1 + z_\text{in}} \right)^{5/2} \! - \frac53 \left( \frac{1 + z}{1 + z_\text{in}} \right) \right] \, , \\ 
\Delta_N P \left( k, z \right) &\simeq \sum_n f_n^2 \frac{6 V}{25 N_n} \, \frac{k^2}{k_{\text{th} (n)}^2 (z_\text{in})} \left[ \frac{D (z)}{D (z_\text{in})} \right]^2  \left[1 - \left( \frac{1 + z}{1 + z_\text{in}} \right)^{5/2} \right]^2 \, . \label{aart}
\end{align}

Individual terms in these sums, in which one sets $f_n = 1$, give the expressions for the differences of partial power spectra for individual components: $\Delta P_n \equiv P_n^\text{warm} - P_n^\text{cold}$.

\section{The velocity field}
\label{sec:field}

Consider first the case of CDM, where the thermal velocities are negligible. In this case, error of simulation will arise due to the coarse-graining of the continuous velocity field. Suppose that we statistically compare two simulations, a fine-grained one and a coarse-grained one, with similar initial conditions.  A qualitative difference between them at every moment of time will be the presence of an additional term $C_\bk^\text{fi}$ in the fine-grained simulation [see Eq.~\eqref{ccc}], which will be absent in the coarse-grained simulation. An important difference between this effect and the effect of thermal velocities will be that, in the present case, perturbation of the initial velocity $\dot \delta_\bk$ can be neglected. Indeed, the time derivative of the density amplitude for a fine-grained system is given by
\begin{equation} \label{velreg}
\dot \delta_\bk = \ri \sum_i \frac{m_i}{M} \bk \dot \bx_i e^{\ri \bk \bx_i} = \ri \sum_I \frac{m_I}{M} e^{\ri \bk \bx_I} \sum_{i_I} \frac{m_{i_I}}{m_I} \bk \left( \dot \bx_I + \dot \by_{i_I} \right) e^{\ri \bk \by_{i_I}} \, .
\end{equation}
For $k \ell \ll 1$, where $\ell$ is a typical comoving separation between the coarse-grained particles placed as $\{ \bx_I \}$, the exponent $e^{\ri \bk \by_{i_I}}$ in \eqref{velreg} can be replaced by unity; then the center-of-mass property
\begin{equation}
\sum_{i_I} \frac{m_{i_I}}{m_I} \by_{i_I} = 0 \, , \qquad \sum_{i_I} \frac{m_{i_I}}{m_I} \dot \by_{i_I} = 0
\end{equation}
will ensure that the fine-grained quantity \eqref{velreg} is approximated by
\begin{equation}
\dot \delta_\bk \approx \ri \sum_I \frac{m_I}{M} \bk \dot \bx_I e^{\ri \bk \bx_I} \, ,
\end{equation}
i.e., is equal to its coarse-grained value with a very high precision.

The effect of $C_\bk^\text{fi}$ can be estimated along the same lines as was done for the case of thermal velocities. Using decomposition \eqref{ccc} and the first expression in \eqref{creg}, one can approximate the evolution of the fine-grained system by the effective equation
\begin{equation} \label{eff-fie}
\ddot \delta_\bk + 2 H \dot \delta_\bk = \left[ 4 \pi G \varrho - \left( \frac{k v_\text{fi}}{a} \right)^2 \right] \delta_\bk + A_\bk - \bar C_\bk \, .
\end{equation}
Here, $v_\text{fi}$ is the characteristic relative velocity between the neighboring particles
in a coarse-grained distribution arising due to inhomogeneity of the velocity
field.\footnote{Note that this is {\em not\/} the total relative velocity between the
particles, part of which constitutes the Hubble velocity, but only the difference between
their peculiar velocities.  It can be defined as $\bv_\text{fi} = a (\dot\bx_i - \dot\bx_j)$,
where $i$ and $j$ label the neighboring particles.} Equation \eqref{eff-fie} effectively
differs from that describing the evolution of the coarse-grained system only by the
presence of the term with $v_\text{fi}$ in the brackets. In this case, we should note that
$v_\text{fi}$ itself depends on the resolution, hence, on the number of particles $N$ in the
coarse-grained system. Therefore, the error of simulation $\Delta P = P_{\rm coarse} -
P_{\rm fine}$ will be determined mainly by the effect that will be characterized
by the wave number $k_\text{fi}$, defined similarly to $k_\text{th}$ but with characteristic
relative field velocity replacing the thermal velocity:
\begin{equation}
k_\text{fi} = \left( \frac{4 \pi G \varrho a^2}{v_\text{fi}^2} \right)^{1/2} \, .
\end{equation}

Returning now to the case of particles with thermal velocities and comparing the two
terms in \eqref{creg}, we can see that whether the thermal velocities are important for
simulations or not is determined by the relation between the characteristic velocities
$v_\text{fi}$ and $v_\text{th}$ or, respectively, $k_\text{fi}$ and $k_\text{th}$.

To estimate the quantities $v_\text{fi}$ and $k_\text{fi} $, we can use the equation that relates
the peculiar velocity field to the density contrast on sub-Hubble spatial scales:
\begin{equation} \label{velo}
\nabla \cdot \bv = \nabla^2 v  = - a \dot \delta = - a H \delta \, ,
\end{equation}
where $v$ is the scalar velocity potential, and we have taken into account that
$\delta \propto a$ at the matter-dominated stage. The difference between peculiar
velocities of particles separated by a small comoving distance $\Delta \bx$ is
\begin{equation} \label{delvel}
\Delta \bv = \left( \Delta \bx \nabla \right) \bv = \left( \Delta \bx \nabla \right) \nabla v \, .
\end{equation}
Using \eqref{velo} and \eqref{delvel}, we calculate the dispersion of the relative
velocity perturbation at this distance:
\begin{equation}
\left\langle \left( \Delta \bv \right)^2 \right\rangle = (a H)^2 \int \frac{d^3 \bk}{(2 \pi)^3} \, \frac{(\bk \Delta \bx)^2}{k^2} P (k) W_k = \frac13 (a H)^2 (\Delta x)^2 \int_0^\infty \frac{d k}{k} \cP (k) W_k \, ,
\end{equation}
where $W_k \geq 0$ is an appropriate window function that cuts the integral at the Nyquist wave number $k = k_N$, and we have used definition \eqref{powerspec} for the dimensionless spectrum $\cP (k)$. In the WDM scenario, where the power spectrum rapidly decays at large wave numbers $k$ [see Eq.~\eqref{wdmpower}, \eqref{transfer}], the last integral can be estimated by the region (in fact, a plateau) where $\cP (k)$ reaches the maximum value $\cP_{\rm max}$, and we have
\begin{equation}
\left\langle \left( \Delta \bv \right)^2 \right\rangle \simeq (a H)^2 (\Delta x)^2 \cP_{\rm max} \, .
\end{equation}

Now, considering the smallest interparticle distance $\Delta x = \left(V / N \right)^{1/3} = \pi / k_N$, we have
\begin{equation}
v_\text{fi}^2 \simeq \pi^2 \left( \frac{a H}{k_N} \right)^2 \cP_{\rm max}
\end{equation}
and
\begin{equation}
k_\text{fi}^2 (z)  \simeq \frac{3 k_N^2}{2 \pi^2 \cP_{\rm max} (z)} \, .
\end{equation}
The quantity $\cP_{\rm max} (z)$ is estimated as
\begin{equation}
\cP_{\rm max} (z) \simeq \frac{\cP_{\rm max} (0)}{(1 + z)^2} \simeq \frac{1}{(1 + z)^2} \, .
\end{equation}
Eventually, we have
\begin{equation} \label{kfi}
k_\text{fi}^2 (z)  \simeq \frac{3}{2 \pi^2} k_N^2 (1 + z)^2 \, .
\end{equation}
For $z \gtrsim 0.5$, the wave number $k_\text{fi}$ exceeds the Nyquist wave number $k_N$. 

The effect of thermal velocities will be unimportant relative to the effect caused by the coarse-graining of the regular velocity field if $k_\text{fi} \lesssim k_\text{th}$. For a thermal relic, the thermal wave number is given by \eqref{kth-th}, and, using \eqref{kfi}, we obtain the estimate of the redshift for which the effect of thermal velocities can be neglected relative to the effect of coarse-graining of the regular velocity field:
\begin{equation} 
1 + z \lesssim 30 \left( \frac{\omega_\text{DM}}{0.12} \right)^{1/9} \left( \frac{m_\text{TR}}{\text{keV}} \right)^{8/9} \left( \frac{L}{50\, \text{Mpc}}\right)^{2/3} \left( \frac{512}{N^{1/3}} \right)^{2/3} \, ,
\end{equation}
where $L = V^{1/3}$ is the comoving size of the simulation box.  For sterile neutrino, using \eqref{kth-sn}, we similarly obtain
\begin{equation} 
1 + z \lesssim 12 \left( \frac{\omega_\text{DM}}{0.12} \right)^{1/3} \left( \frac{m_\text{SN}}{\text{keV}} \right)^{2/3} \left( \frac{L}{50\, \text{Mpc}}\right)^{2/3} \left( \frac{512}{N^{1/3}} \right)^{2/3} \, .
\end{equation}
Our crude estimates show that, for dark-matter particle masses in the keV range, switching thermal velocities in $N$-body simulations is relevant only for sufficiently high redshifts.

\section{Discussion}
\label{sec:discuss}

In this paper, we investigated analytically the expected effects of thermal velocities in $N$-body simulations with warm dark matter.  In numerical simulations, thermal velocities of WDM particles are taken into account by adding random initial velocities to the simulation particles according to the velocity distribution function of WDM calculated at the initial moment of time.  However, because of obvious computational limitations, a huge number of DM particles are represented as one body in $N$-body simulations. Formally, the average thermal velocity of such a collection of particles is very close to zero, which raises the issue of correctness of the procedure of adding thermal velocities to the simulation particles.  By presenting the evolution equation for the density contrast in the form \eqref{nonlin} due to Peebles \cite{Peebles-AA, Peebles:1980yev}, we were able to see that this procedure is legitimate as long as it produces the same velocity term $C_\bk$.  We have elaborated on this in Sec.~\ref{sec:coarse}.

The physical effect of thermal velocities, surviving in the limit of $N \to \infty$, is described by Eq.~\eqref{phys} on spatial scales much larger than the thermal scale of the warm dark mater ($k \ll k_\text{th}$) defined in \eqref{kth}. It causes small suppression of the power spectrum and is understandable as the effect of free-streaming. Its quantitative features are very similar to the effect of suppression of Jeans instability by pressure and the related speed of sound. The thermal spatial scale $2 \pi / k_\text{th}$ is a close analog of the Jeans scale and is determined by the thermal velocity distribution.  Sometimes it is also called the free-streaming scale \cite{Boyarsky:2008xj}.  

Along with the physical effect, $N$-body simulations contain artificial effects connected with discreteness.  We have shown that the dominating artificial effect of switching thermal velocities consists in the perturbation \eqref{noise} of the initial time derivative $\dot \delta_\bk$ of the density profile.  Its evolution eventually produces an additional term \eqref{art} in the power spectrum.  Unlike the physical effect of thermal velocities, this effect does not depend on the original power spectrum (it is additive rather than multiplicative), but depends on the particle number density $N/V$ in the simulations.  We have obtained a simple analytic formula \eqref{main2} combining all effects of thermal velocities in the power spectrum as well as the usual shot noise.  In the case of large initial redshift, it is further simplified to \eqref{main1}.  As a specific discreteness effect caused by thermal velocities in simulations, the model predicts a turnover in the behavior of the power spectrum at certain wave number $k_*$, which is estimated implicitly by formula \eqref{k*}. The power spectrum at $k > k_*$ starts growing artificially as $P_\text{warm} (k) \propto k^2$, corresponding to the last term in \eqref{main1}. In real simulations, this is likely to cause the production of a large number of spurious halos on these scales, similar to those observed in simulations with initially ``cold'' particles \cite{Wang:2007he}.

Perhaps, the noise in the initial condition \eqref{noise} can be reduced if one adds thermal velocities not randomly but with correlations among neighbors in such a way that they average to zero over certain small clusters of simulation particles.  For instance, after choosing randomly a thermal velocity for the first particle, one could assign exactly opposite thermal velocity to its randomly chosen neighbor, and then repeat this procedure with other particle pairs. One could also do the same with triples of particles, and so on. Whether such scheme is going to work requires special simulations.

We generalized our model of physical and discreteness effects of thermal velocities to a dark-matter system consisting of several components in Sec.~\ref{sec:several}.  Equations \eqref{aphys} and \eqref{aart} describe, respectively, the physical and discreteness effects on the total power spectrum, with individual terms in the sums describing the effects on the power spectra for each dark-matter component.

Effective coarse-graining of the regular velocity field in $N$-body simulations is a source of another discreteness effect in the evolution of the power spectrum. In Sec.~\ref{sec:field}, we show that it also tends to enhance the power spectrum because of the presence of the term proportional to $k^2 / k_\text{fi}^2$ in the effective evolution equation for the density contrast, where $k_\text{fi}$ is the corresponding characteristic wave number connected with the coarse-graining of the smooth velocity field.  For a given particle number $N$, we estimate the redshift below which this effect exceeds the effect of thermal velocities, in which case the latter can be neglected in simulations.

According to our results, whether is is useful to switch on random thermal velocities in $N$-body simulations depends on the desired precision in the power spectrum at relatively small wave numbers, where the effect of thermal velocities can be captured with negligible numerical artifacts. In cases where such numerical artifacts are unacceptably large, it may be advisable just to make the theoretical correction for the spectrum that uses cold particles, as given by Eq.~\eqref{phys}. 

To capture the effects of thermal velocities under the conditions of coarse-graining naturally arising in $N$-body simulations, we used certain model assumptions that are difficult to control on a rigorous basis. For example, we neglected (with some justification) possible correlations between regular and thermal velocities arising in the course of evolution, or contributions of nonlinear and regular velocity terms, $A_k$ and $C_k^\text{reg}$, in the basic equation \eqref{exact}. However, comparison of our results for the turnover scale $k_\star$, determined by \eqref{kvel}, with the available results of numerical simulations of warm dark matter \cite{Leo:2017zff} shows that our model works quite well in this respect.

It would be interesting to extend the present model of the effects of thermal velocities to the stage of fully nonlinear evolution of the density profile \cite{Leo:2017zff} and to more sophisticated computational algorithms, and we hope to return to these issues in the future.

\section*{Acknowledgments}

We are grateful to Alexey Boyarsky and Oleg Ruchayskiy for valuable discussion and to the authors of paper \cite{Leo:2017zff} for the consent to reproduce their figure. We also thank the anonymous referees for their valuable comments. This work was supported by the National Research Foundation of Ukraine under Project No.~2020.02/0073.  Y.\,S.\@ acknowledges support from the Simons Foundation.

\appendix

\section{\uppercase{Definition of the power spectrum}}
\label{app:power}

\subsection{Infinite space}

In the infinite space, we define $\rho (\bx)$ to be the matter density (at a given moment
of time; we suppress the time coordinate), so that
\begin{equation}
a^3 \int_V \rho ( \bx ) d^3 \bx = M_V
\end{equation}
is the total mass in a {\em comoving\/} volume $V$.  For an infinite space, the quantity
\begin{equation}
\lim_{V \to \infty} \frac{M_V}{a^3 V} = \varrho
\end{equation}
is the background density, and the dimensionless density fluctuation is defined as
\begin{equation}
\delta( \bx) = \frac{\rho (\bx) - \varrho}{\varrho} = \frac{\rho (\bx)}{\varrho} - 1 \, .
\end{equation}

We define the Fourier transform in the infinite comoving space as
\begin{equation} \label{fourier}
\delta_\bk = \int \delta (\bx) e^{\ri \bk \bx} d^3 \bx \, , \qquad \delta ( \bx) = \int \delta_\bk e^{- \ri \bk \bx} \frac{d^3 \bk}{(2 \pi)^3}  \, , \qquad \delta_\bk^* = \delta_{- \bk}^{} \, .
\end{equation}

For small density contrast, the density fluctuation is assumed to be a homogeneous
Gaussian process with correlation function
\begin{equation} \label{corrfunc}
\xi(x) = \left\langle \delta(0) \delta(\bx) \right\rangle = \int P (k) e^{- \ri \bk \bx} \frac{d^3 \bk}{(2 \pi)^3} = \int \cP (k) \frac{\sin k x}{k x} \frac{d k}{k} \, ,
\end{equation}
where the dimensionless quantity
\begin{equation} \label{powerspec}
\cP (k) \equiv \frac{k^3}{2 \pi^2} P (k)
\end{equation}
determines the rms amplitude $\sigma_R$ of mass fluctuation on the comoving scale $R$ via
\begin{equation}
\sigma_R^2 \equiv \left\langle \left( \frac{\delta M}{M} \right)^2_R \right\rangle = \int W_R (k) \cP(k) \frac{ d k }{k} \, ,
\end{equation}
where $W_R (k)$ is an appropriate window function.

Equation \eqref{corrfunc} together with \eqref{fourier} implies
\begin{equation}
\left\langle \delta_\bk^{} \delta_\bp^* \right\rangle = P (k) (2 \pi)^3 \delta_\text{D} ( \bk - \bp ) \, ,
\end{equation}
where $\delta_\text{D} (\bk)$ is Dirac's delta-function.

\subsection{Finite space}

Simulations are done in a finite volume; therefore, we must introduce the counterparts of
the preceding quantities in this case.  The space integrals are now performed over the
total finite comoving volume $V$, and the Fourier space becomes discrete, with the 
orthogonality property 
\begin{equation}
\int_V e^{\ri \left( \bk - \bk' \right) \bx } d^3 \bx = V \delta_{\bk, \bk'} \, ,
\end{equation}
where $\delta_{\bk, \bk'}$ is the Kronecker symbol. The background density is now
\begin{equation}
\varrho = \frac{M}{a^3 V} \, ,
\end{equation}
where $M$ is the total mass.  The Fourier transform is defined similarly to
\eqref{fourier} with normalization over the total volume:
\begin{equation} \label{d-fourier}
\delta_\bk = \int \delta (\bx) e^{\ri \bk \bx} \frac{d^3 \bx}{V} \, , \qquad \delta ( \bx) = \sum_\bk \delta_\bk e^{- \ri \bk \bx} \, , \qquad \delta_\bk^* = \delta_{- \bk}^{} \, .
\end{equation}
With this convention, the quantity $\delta_\bk$ is dimensionless, as is $\delta (\bx)$.

In a large finite volume, the sum over momenta is approximated by an integral as follows:
\begin{equation}
\sum_\bk \to \int \frac{ V d^3 \bk}{(2 \pi)^3} \, .
\end{equation}
Therefore, we have the following connection between the Fourier amplitudes and power
spectrum in the case of finite volume:
\begin{equation}
V \left\langle \delta_\bk^{} \delta_\bp^* \right\rangle = P (k) \delta_{\bk,\bp}
\, .
\end{equation}
In particular,
\begin{equation} \label{normalization}
V \left\langle \left| \delta_\bk \right|^2 \right\rangle = P(k) = \frac{ 2 \pi^2 \cP (k)}{k^3} \, .
\end{equation}
This gives normalization of the rms amplitude of the Fourier coefficients $\delta_\bk$ to
be used in the estimates, with power spectrum defined in \eqref{corrfunc} and \eqref{powerspec}.

\section{\uppercase{Power spectra}}
\label{app:spectra}

\subsection{Cold dark matter}

The {\em today's\/} linear power spectrum in the $\Lambda$CDM model can be approximated as \cite{Efstathiou:1992sy}:
\begin{equation}
P_\text{CDM} (k) = \frac{B k}{\left( 1 + \left[ \alpha k + ( \beta k )^{3/2} + (\gamma k)^2 \right]^\nu \right)^{2/\nu}} \, ,
\end{equation}
where $\alpha = (6.4/\omega_\text{m})$~Mpc, $\beta = (3.0/\omega_\text{m}) $~Mpc, $\gamma =
(1.7/\omega_\text{m})$~Mpc, $\nu = 1.13$, and $\omega_\text{m} = \Omega_\text{m} h^2$. The comoving quantity
$k$ is measured in Mpc$^{-1}$. In accordance with the CMB power-spectrum normalization, we have
\begin{equation}
B = \frac{8 \pi^2 g_0^2 A c^4}{25 \Omega_\text{m}^2 H_0^4} \, ,
\end{equation}
where the scalar-type perturbation amplitude \cite{Planck:2018vyg} $A = 2.1 \times 10^{-9}$, and $g_0 \approx 0.75$.  Therefore,
\begin{equation}
B \approx 1.5 \times 10^7 \left(\frac{0.143}{\omega_\text{m}} \right)^2 \, \text{Mpc}^4 \, .
\end{equation}

The linear power spectrum at arbitrary $z$ is then calculated approximately as
\begin{equation}
P_\text{CDM} (k, z) \approx \frac{g^2 (z) P_\text{CDM} (k)}{(1 + z)^2} \, ,
\end{equation}
where
\begin{equation}
g (z) = \left[ 1 + \frac{\Omega_\Lambda}{\Omega_\text{m}} \right]^{1/3} \left[ 1 +
\frac{\Omega_\Lambda}{\Omega_\text{m} (1 + z)^3} \right]^{- 1/3}
\end{equation}
is the factor taking into account the presence of the cosmological constant.

\subsection{Warm dark matter}

In the linear theory with WDM, the today's spectrum $P_\text{WDM} (k)$ can be related to that of the CDM
$P_\text{CDM} (k)$ by a transfer function $T(k)$:
\begin{equation} \label{wdmpower}
P_\text{WDM} (k) =  T^2 (k)\, P_\text{CDM} (k) \, .
\end{equation}
In the case of thermal relics with phase-space distribution \eqref{fermi}, the transfer function is approximated by \cite{Viel:2005qj}
\begin{equation} \label{transfer}
T (k) = \left[ 1 + \left( \frac{k}{k_0} \right)^{2 \alpha} \right]^{-5/\alpha} \, ,
\end{equation}
where $\alpha = 1.12$, and
\begin{equation}\label{cut-th}
k_0 \approx 14.3 \left(\frac{m_\text{TR}}{\text{keV}} \right)^{1.11} \left(\frac{0.25 \times
	0.7^2}{\Omega_\text{DM} h^2} \right)^{0.11} \, \text{Mpc}^{-1}
\end{equation}
is the characteristic free-streaming scale, with $m_\text{TR}$ being the mass of the thermal
relic. The equivalent mass $m_\text{SN}$ of sterile neutrino with spectrum \eqref{fermi} and
temperature $T_\text{DM} = T_\nu$ that produces the same transfer function is related to the
mass of thermal relic $m_\text{TR}$ via \cite[Eq.~(5)]{Viel:2005qj}
\begin{equation}
\frac{m_\text{SN}}{\text{keV}} = 4.43 \left( \frac{m_\text{TR}}{\text{keV}} \right)^{4/3} \left(\frac{0.25 \times 0.7^2}{\Omega_\text{DM} h^2} \right)^{1/3} \, .
\end{equation}

Expressing $m_\text{TR}$ through $m_\text{SN}$, we then get
\begin{equation}
\frac{m_\text{TR}}{\text{keV}} \approx 0.33 \left(\frac{m_\text{SN}}{\text{keV}}\right)^{3/4} \left(
\frac{\Omega_\text{DM} h^2}{0.25 \times 0.7^2} \right)^{1/4} \, .
\end{equation}
In the case of sterile neutrino, we will have
\begin{equation} \label{cut-sn}
k_0 \approx 4.2 \left( \frac{m_\text{SN}}{\text{keV}} \right)^{0.83} \left( \frac{\Omega_\text{DM} h^2}{0.25 \times 0.7^2} \right)^{0.17} \, \text{Mpc}^{-1} \, .
\end{equation}

One can observe that the cutoff wave numbers \eqref{cut-th} and \eqref{cut-sn} are much smaller than the typical corresponding thermal wave numbers \eqref{kth-th} and \eqref{kth-sn}.

\bibliography{wdm.bib}

\end{document}